\documentclass[a4paper,11pt]{article}

\usepackage{jinstpub} 

\usepackage{lineno}
\title{\boldmath Implementation of ACTS for STCF track reconstruction}

\author[a]{Xiaocong Ai,}
\author[b]{Xingtao Huang,}
\author[a]{Yi Liu\note{Corresponding author}}

\affiliation[a]{School of Physics and Microelectronics, Zhengzhou University, Zhengzhou, Henan, 450001, China}
\affiliation[b]{Key Laboratory of Particle Physics and Particle Irradiation (MOE), Institute of Frontier and Interdisciplinary Science, Shandong University, Qingdao, Shandong, 266327, China}

\emailAdd{yiliu@zzu.edu.cn}

\abstract{With an electron-positron collider operating at center-of-mass-energy 2-7 GeV and a peak luminosity above $0.5\times10^{35} cm^{-2} s^{-1}$, the STCF physics program will provide an unique platform for in-depth studies of hadron structure and non-perturbative strong interaction as well as probing new physics beyond the Standard Model in the $\tau$-Charm sector, succeeding the present Beijing Electron-Positron Collider. To fulfill the physics targets and further maximize the physics potential at STCF, the STCF tracking software should have capability to reconstruct charged particles with high efficiency and excellent momentum resolution, especially for the charged particles with low transverse momentum down to 50 MeV.
A Common Tracking Software (ACTS) providing a set of detector-independent tracking algorithms is adopted for reconstructing charged tracks with the information of two sub-detectors, a $\mu$RWELL-based
inner tracker and a drift chamber, at STCF. This is the first demonstration of ACTS for a drift chamber. The implementation details and performance of track reconstruction are presented.
}


\begin{document}
\maketitle
\flushbottom

\section{Introduction}
\label{sec:intro}
Hadron physics plays an important role in studying Quantum Chromodynamics (QCD) in the low energy region, where perturbative QCD is not applicable due to color confinement. A multi-GeV $e^+e^-$ collider operating in the $\tau$-charm sector provides an unique platform for studying non-perturbative QCD and strong interactions of the Standard Model (SM). Currently, the Beijing Electron Positron Collider (BEPCII)- Beijing Spectrometer (BESIII)~\cite{ABLIKIM2010345} is the only facility at such energy region in the world. With a luminosity of two-orders higher (peak luminosity is above $0.5\times10^{35} cm^{-2} s^{-1}$) and the energy region (2-7 GeV) wider than those at BEPCII, the future $\tau$-charm factory, Super Tau-Charm Facility (STCF)~\cite{Luo:IPAC2019-MOPRB031}, aims to continue and extend the physics programs at BESIII in the post-BEPCII era. The physics goals of STCF include in-depth studies of hadron structure and the nature of non-perturbative strong interactions, exploring the asymmetry of matter-antimatter and searches for particles and physics beyond the SM.

To fullfill the physics goals of STCF, the charged tracks must be reconstructed with both high efficiency and high precision, as the capability of charged tracks reconstruction, i.e.~tracking, has significant impact on the performance of vertex reconstruction, particle identification and background suppression. In particular, among the final state particles of many important processes for studying CP Violation, CKM elements, $D^0-\bar D^0$ mixing and so on at STCF, there are a considerable number of particles with momentum lower than 400 MeV. Therefore, good tracking efficiency for particles with momentum below a few hundreds of MeV down to 50 MeV is very important and also very challenging.

A Common Tracking Software (ACTS)~\cite{Ai2022,Salzburger_A_Common_Tracking_2021} is a tracking toolkit with a set of detector-independent and framework-independent modular tools dedicated to track reconstruction and vertex reconstruction for High Energy Physics (HEP) experiments. It has been used by a variety of HEP experiments, e.g. ATLAS~\cite{ATL-PHYS-PUB-2021-012} and sPHENIX~\cite{Osborn2021}. However, its application on $e^+e^-$ colliders at the precision frontier is very limited so far. In particular, ACTS has not been used for track reconstruction with a gaseous drift chamber yet.

In this study, the tracking performance of STCF with a fully gaseous tracking system consisting of a $\mu$RWELL~\cite{Bencivenni_2017}-based inner tracker and a drift chamber is studied using the Kalman Filter~\cite{kalman1960} based tracking algorithms of ACTS. The manuscript is organized as follows. In Section~\ref{sec:stcf}, the STCF detector is introduced. The STCF offline software framework and ACTS are described in Section~\ref{sec:software}. Section~\ref{sec:reco} focuses on implementation of ACTS for track reconstruction and the performance is presented in Section~\ref{sec:perf}. A brief conclusion is given in Section~\ref{sec:conclusion}.
\section{The STCF detector}
\label{sec:stcf}

\begin{figure}[htbp]
\begin{center}
\includegraphics[width=0.7\textwidth]{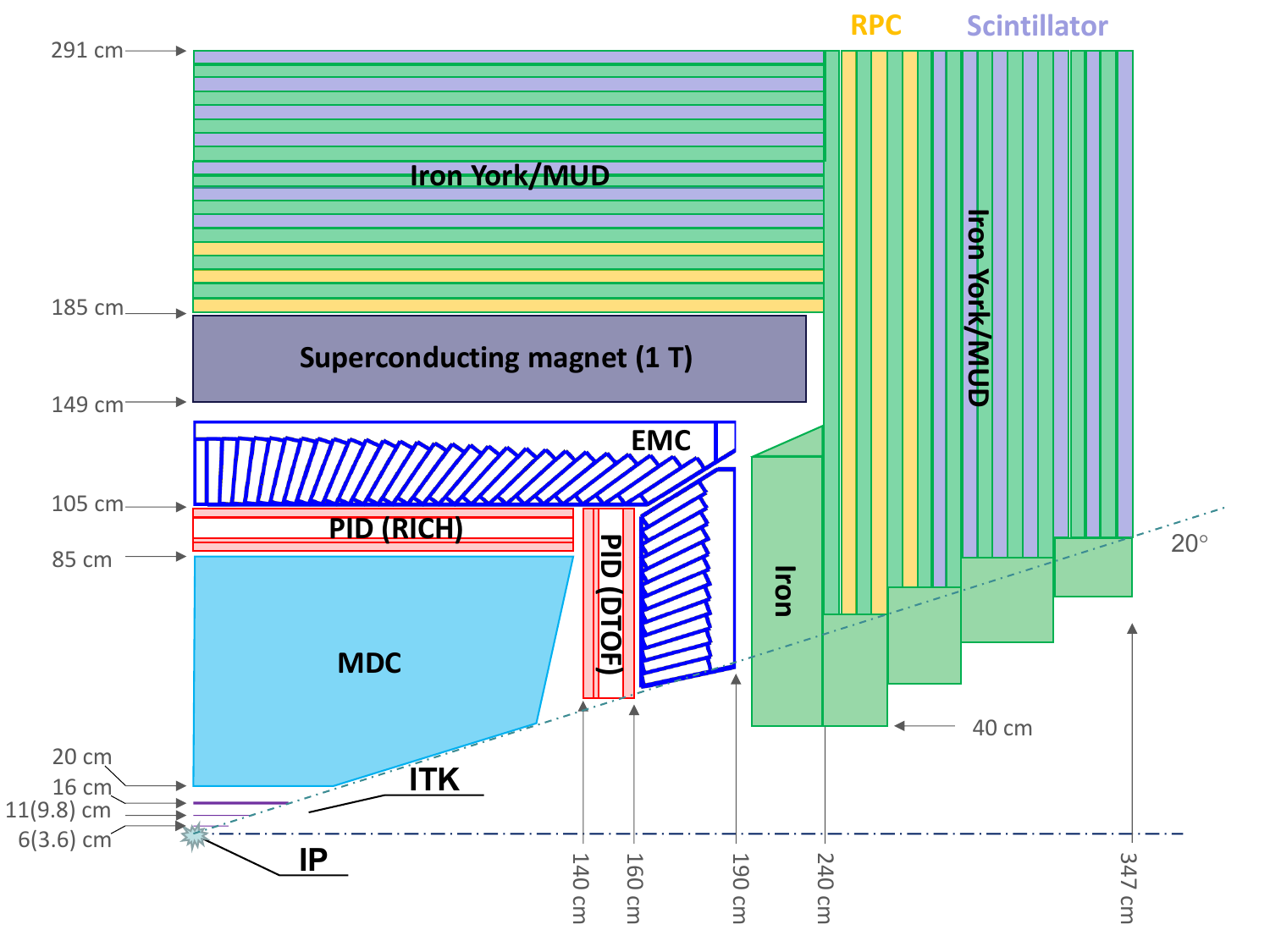}
\caption{Schematic layout of the STCF detector. The number in brackets indicate the radii of the MAPS-based ITK.}
\label{fig:fulldetector_overview}
\end{center}
\end{figure}

The STCF detector is designed to provide a coverage of almost the entire solid angle around the collision point. The baseline layout of the STCF detector is shown in Figure~\ref{fig:fulldetector_overview}. It consists of a tracking system which includes an inner tracker (ITK) and a Main Drift Chamber (MDC), a Ring Imaging Cherenkov (RICH) detector and a DIRC~\cite{ADAM2005281}-like time-of-flight (DTOF) detector for particle identification in the barrel and endcap, respectively, a homogeneous Electro-magnetic Calorimeter (EMC), a superconducting solenoid magnetic producing a 1 Tesla axial magnetic field, and a Muon Detector (MUD) at the outermost of the detector system.

The tracking system provides position measurement for charged particles in the range $|\textrm{cos}\theta|<0.94$. To achieve the physics goals of STCF, it's required to provide a tracking efficiency above 99\% (90\%) for charged tracks with $p_T >$ 300 (100) MeV, and a momentum resolution of $\sigma_{p_T}/{p_T}<0.5$\% for charged tracks with $p_T = 1$ GeV.
There are two options for the ITK, the MAPS-based ITK and $\mu$RWELL-based ITK. This study is based on the $\mu$RWELL-based ITK, which consists of three layers of light-material $\mu$RWELL-based gaseous detectors around the beam pipe with the inner radii of 60 mm, 110 mm and 160 mm, respectively. 
It provides a spatial resolution around 100 $\mu$m in the $r$-$\phi$ direction and around 400 $\mu$m in the $z$ direction. The MDC adopts a square cell and a superlayer wire configuration, and uses He/C$_{3}$H$_{8}$~(60/40) as the working gas. There are eight superlayers and each superlayer contains six layers of drift cells. The superlayers alternate between axial ("A") orientation, aligned with the direction of the beam line, and stereo ("U", "V") orientation. The eight superlayers are arranged in AUVAUVAA. The inner radius and outer radius of MDC is 200 mm and 850 mm, respectively. The MDC is designed to provide a spatial resolution between 120 $\mu$m and 130 $\mu$m, and a dE/dx resolution around 6\%.

By measuring the characteristic radiation angle or spatial-time hit pattern of Cherenkov photons, the RICH and DTOF are designed to facilitate identification of charged hadrons ($\pi$, $K$, proton) with momentum from 700 MeV to 2 GeV. The scintillating crystals based EMC provides measurement of energy and direction for photons as well as identification for photons, electrons and charged hadrons. The MUD consists of both plastic scintillator strips and resistive plate chambers to provide information for identification between $\pi$ and $\mu$.
\section{The STCF software framework and ACTS}
\label{sec:software}

The Offline Software System of Super Tau-Charm Facility (OSCAR)~\cite{oscar} is the offline event processing framework for STCF. It consists of an interface for external third-party software, a framework providing common functionalities for data processing and a set of application tools for event generation, simulation, reconstruction and physics analysis. 
The physics generators with high precision, e.g. KKMC~\cite{kkmc}, EVTGEN~\cite{evtgen}, are integrated into OSCAR for simulating the $\tau$-charm physics processes.
The Detector Description Toolkit, DD4Hep~\cite{dd4hep}, is adopted to describe STCF detector geometry with all geometric parameters stored in the compact files with eXtensible Markup Language (XML)~\cite{xml}. Geant4~\cite{AGOSTINELLI2003250} is integrated into OSCAR for full simulation of the interaction of particles with detector.

To make a common tracking toolkit, the geometry and navigation model of ACTS, and its Event Data for describing measurements, track parameters and vertex parameters, are designed to be independent on the details of specific experiments. ACTS is written in modern C++17 and features stateless modules to facilitate multi-threaded event reconstruction, in compliance with modern multi-core CPU architectures.

In ACTS, the track parameters can be described with either bound (also called local) track parameters or free (also called global) track parameters. With the bound track parameters, the position of the track is represented in local coordinates of a detector surface, and the track direction is described by azimuthal and polar angles. With the free track parameters, the position is represented in global coordinates of the detector and the track direction is represented by a unit vector. Both the bound track parameters and free track parameters include a curvature parameter and the flight time of the particle.


The bound track parameters are represented as:
\begin{equation}
b = (l_0, l_1, \phi, \theta, q/p, t),
\end{equation}
where the first two track parameters are the coordinates of the track in the local coordinates of a reference surface, $\phi$ and $\theta$ are the azimuthal and polar angles of the track direction, $q/p$ is the ratio of charge $q$ and momentum $p$, and $t$ is time coordinate of a particle in space-time.

The measurement is designed to be a subset of the bound track parameters. Therefore, projection from the bound track parameters to the measurement is realized using a projection matrix.

Depending on the detector readout geometry, surface of dedicated type is used as the reference surface of the bound track parameters and measurement. For STCF, the cylinder surface and line surface are used for the ITK and the MDC, respectively.
The line surface in ACTS is also used for representing the track parameters near the interaction point, i.e.~perigee track parameters. In such case, the $l_0$ and $l_1$ represent the transverse impact parameter $d_0$ and longitudinal impact track parameter $z_0$, respectively.

\section{STCF track reconstruction using ACTS}
\label{sec:reco}

\begin{figure}[htbp]
\begin{center}
\includegraphics[width=0.8\textwidth]{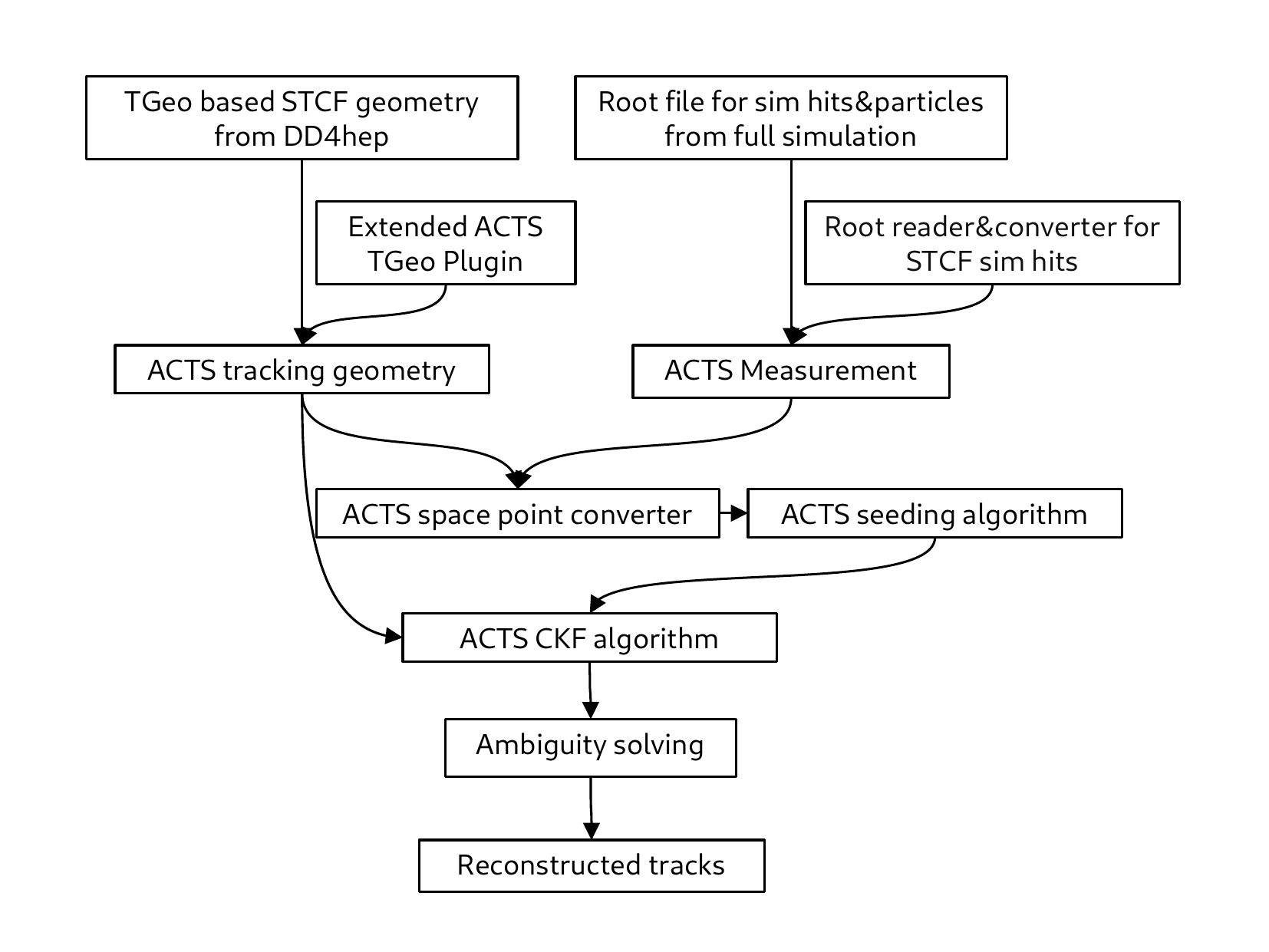}
\caption{The workflow of applying ACTS for track reconstruction at STCF.}
\label{fig:reco_chain}
\end{center}
\end{figure}

Figure~\ref{fig:reco_chain} shows the workflow of applying the ACTS tracking toolkit for track reconstruction at STCF. The interface of ACTS with experiments is extended for STCF. The Combinatorial Kalman Filter (CKF)~\cite{Fruhwirth2021,Braun:1540,BERTACCHI2021107610} implemented in ACTS is used to find the tracks based on the seeds provided by the ACTS seed finding algorithm, and then it is followed by a subsequent ambiguity solving step to remove incomplete or duplicate tracks.

\subsection{Extension of interface for STCF}
Based on the interfaces provided by ACTS, several geometry plugins are developed to facilitate transformation of an experiment's geometry in an existing representation, e.g. DD4hep or TGeo~\cite{BRUN2003676}, into an internal geometry description of ACTS. The ACTS TGeo plugin is extended for both ITK and MDC. For ITK, the tube containing the signal readout unit in each $\mu$RWELL layer is transformed into a sensitive cylinder surface. For MDC, each sense wire in a drift cell is transformed into a line surface. The dedicated material mapping tools~\cite{Ai2022} in ACTS are used to project the detailed material description into internal auxiliary surfaces of ACTS geometry.

A ROOT~\cite{Brun:1997pa}-based reader is implemented to read the simulated hits from full simulation and convert them into ACTS measurements after taking into account the resolution of detectors. The ITK measurement is two dimensional, and the $l_0$ and $l_1$ represent $r\cdot \varphi$ and $l_z$, respectively, with $r$ being the radius of the cylinder, $\varphi$ being the azimuthal angle of the track position on the local $x$-$y$ plane of the cylinder frame, and $l_z$ being the local $z$ coordinate of the track position in the cylinder frame. The MDC measurement is one dimensional with $l_0$ representing the drift distance of the ionized electrons to the anode wire of the MDC cell.

Details of the interface extension can be found in Ref.~\cite{oscar_manual}.

\subsection{Seed finding}
Track seeds, i.e.~triplets of measurements from increasing radii which are likely to belong to the same track, are created from the measurements of the ITK detector using the ACTS track seed finding algorithm. Using the coordinates of the measurements in global coordinate frame, the algorithm groups the measurements based on a number of criteria related with the fiducial coverage of the detector and the present magnetic field. Under the assumption of a homogeneous magnetic field along global $z$ axis, a track follows a helix trajectory, i.e.~a circle on the $x$-$y$ plane and a straight line in the $s$\footnote{The $s$ is the path length of the track on the $x-y$ plane}-$z$ plane.
For each seed candidate, the curvature and center of the circle on the $x$-$y$ plane can be obtained using Conformal transform~\cite{Fruhwirth2021}, and then used to derive the transverse impact parameter on the $x$-$y$ plane. 
The seeds are filtered by requiring estimated quantities of the seeds, such as curvatures, transverse impact parameters and polar angles, to pass individual criteria, as detailed in Ref.~\cite{garg2023exploration}. The criteria are designed to be configurable parameters of the ACTS track seed finding algorithm and optimized using the Optuna Hyperparameter optimization~\cite{optuna} approach, which has been implemented into ACTS, with the following objective function (score):
\begin{equation}
\textrm{Score} = \textrm{Efficiency} - (\textrm{Fake rate} + \frac{\textrm{Duplicate rate}}{k}), k = 7,
\label{eq:score}
\end{equation}
where $k$ is a weighting parameter tuned for optimal trade-off between the efficiency of track finding and the rate of fake or duplicate tracks.

\subsection{Simultaneous track finding and track fitting using CKF}
\begin{figure}[htbp]
\centering
\includegraphics[width=.6\textwidth]{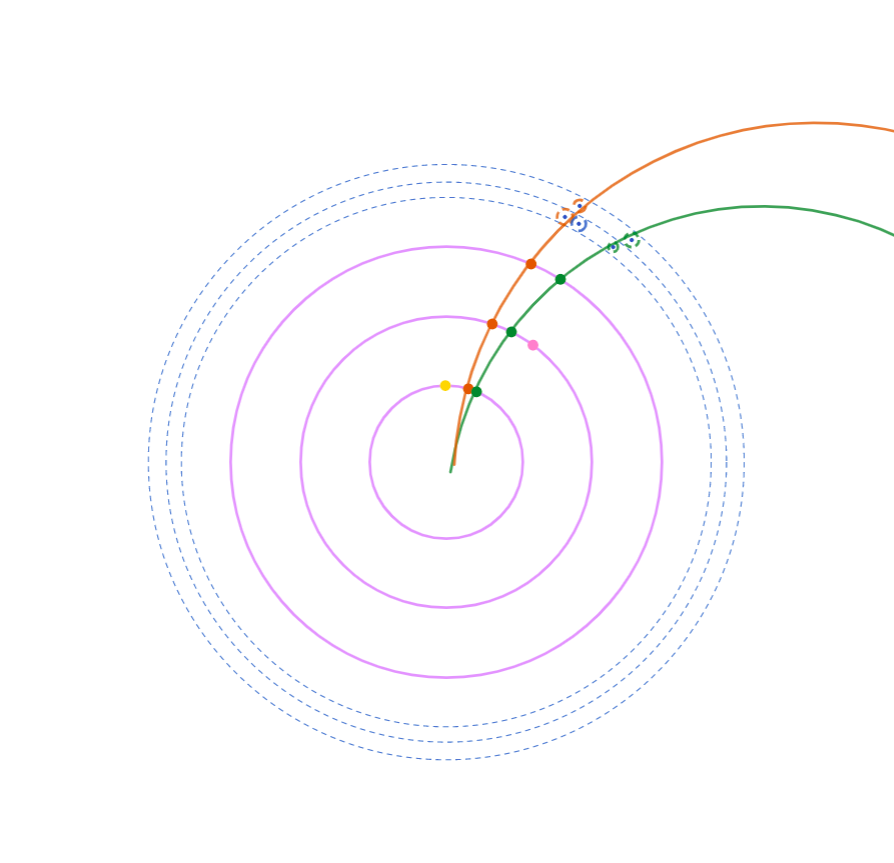}
\caption{\label{fig:ckf} Illustration of track finding using CKF with ITK (purple) and MDC (blue) of STCF. Only two MDC layers are shown here.}
\end{figure}

The CKF performs track finding through track fitting. Starting from track parameters estimated from a seed, it searches and associates compatible measurements to the track iteratively during track propagation, as shown in Fig.~\ref{fig:ckf}. When multiple measurements are found upon a propagation step, a prediction $\chi^2$ is calculated using the predicted track parameters and the measurement for each measurement candidate,
\begin{equation}
\chi^2 = r^T(HCH^T + V)^{-1}r,
\end{equation}
where $H$ is the projection matrix from the bound track parameters to the measurement, $C$ is the covariance of the bound track parameters, $V$ is the covariance of the measurement and $r$ is the residual,
\begin{equation}
r = m - Hb,
\end{equation}
where $m$ and $b$ are the measurement vector and bound track parameters vector, respectively. Those measurements with $\chi^2$ below a threshold, $\chi^2_{\textrm{min}}$, are considered to compatible to the track and sorted based on $\chi^2$ in ascending order. In the case of more than one compatible measurements, i.e.~$n^{\textrm{meas}}>1$, the propagation is split into $n_{\textrm{max}}^{\textrm{branch}}$ branches using the $n_{\textrm{max}}^{\textrm{branch}}$ sets of filtered track parameters based on the first $n_{\textrm{max}}^{\textrm{branch}}$ measurements among the sorted measurements. The $\chi^2_{\textrm{min}}$ and $n_{\textrm{max}}^{\textrm{branch}}$ (reduced to $n^{\textrm{meas}}$ if $n_{\textrm{max}}^{\textrm{branch}} > n^{\textrm{meas}}$) are configurable parameters of the ACTS CKF algorithm and optimized based on the objective function in Eq.~\ref{eq:score}. In the study presented in this paper, the optimal criterion for $n_{\textrm{max}}^{\textrm{branch}}$ is found to be $n_{\textrm{max}}^{\textrm{branch}} = 1$, which corresponds to a sequential Kalman Filter with only the most compatible measurement associated to the track at each propagation.


Upon the ending of forward track propagation, the Kalman smoothing of the track parameters is performed for each found track candidate. The perigee track parameters for each track candidate is obtained by propagating the smoothed track parameters at the first measurement plane to the perigee plane.

For a drift chamber, there is an inherent left-right ambiguity of the drift distance. Such ambiguity can be naturally resolved by a global track finding algorithm, such as the Legendre algorithm~\cite{BERTACCHI2021107610} and Hough Transform~\cite{Duda1972UseOT} used by the Belle II experiment~\cite{BERTACCHI2021107610,Braun:1540}. However, it's more difficult to resolve in a local track finding algorithm such as CKF. Simply splitting the propagation into two branches for both signs of a drift distance in CKF, if both signs are compatible with the prediction, leads to penalties from much increased combinatorics, duplicate tracks and CPU time. In this study, the left-right sign of the drift distance for a MDC measurement is taken to be the same as that of the predicted track parameters, i.e.~the sign which provides a smaller $\chi^2$ is chosen. In a future study, resolving the left-right ambiguity in CKF using the Deterministic Annealing Filter~\cite{FRUHWIRTH1999197} which remains to be implemented into ACTS will be explored.

\subsection{Ambiguity solving}
A threshold is set on the minimum number of measurements associated to a track to remove incomplete tracks. Because the charged track multiplicity is low at STCF and beam induced background is not considered in this study, the fake tracks among the found track candidates, i.e.~tracks cannot be associated to any simulated particle, are negligible.

However, it can happen that multiple tracks are associated to the same simulated particle, i.e.~duplicate tracks are present. This is mainly due to the presence of duplicate seeds, in particular in events with low momentum looping tracks. If two track candidates have at least two shared measurements, the one with less measurements is removed from the track candidates. This procedure is repeated iteratively until any two of the track candidates have no more than one shared measurement.

\section{Track reconstruction performance}
\label{sec:perf}
The track reconstruction performance of the STCF tracking system following the track reconstruction procedure in Section~\ref{sec:reco} is discussed in this section.

\subsection{Monte-Carlo samples}
\begin{figure}[htbp]
\centering
\includegraphics[width=.45\textwidth]{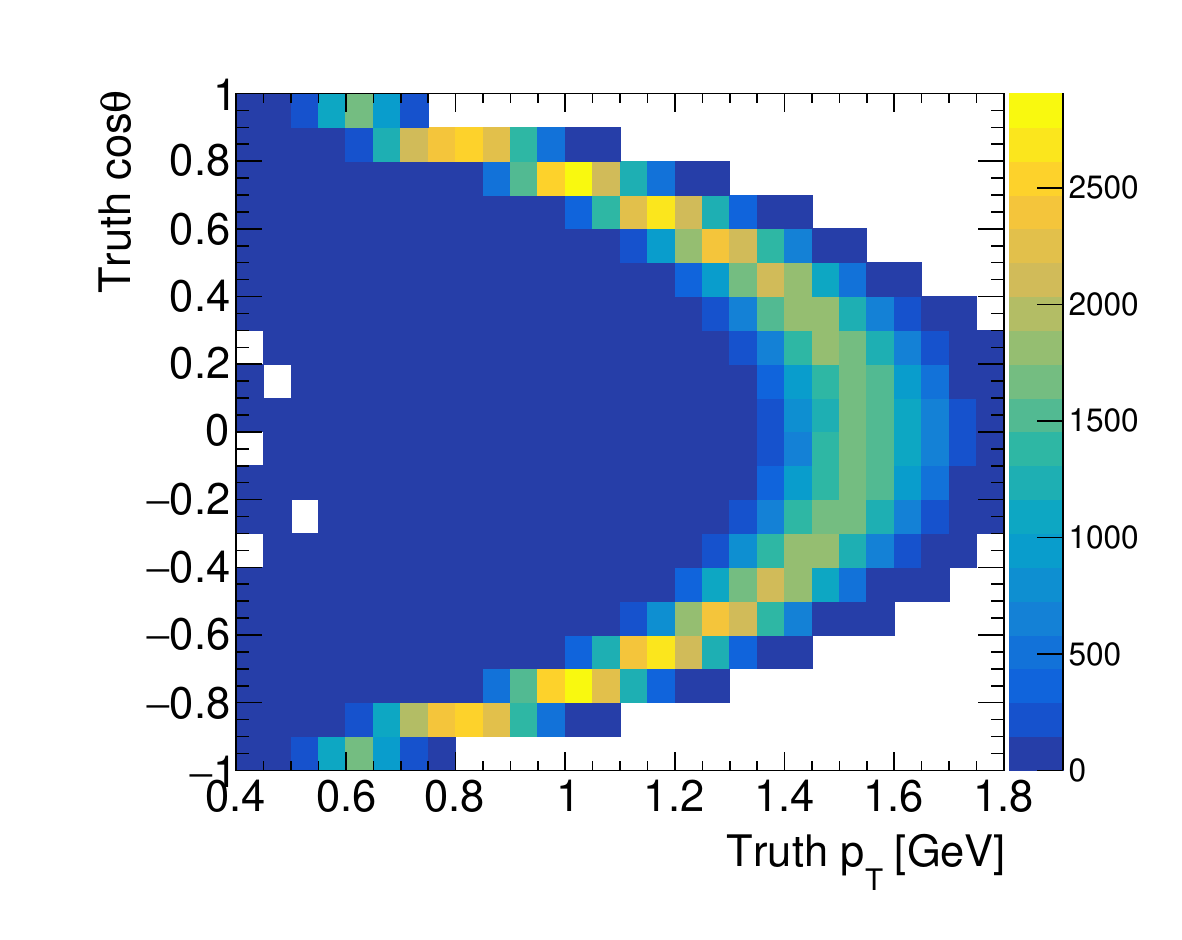}
\includegraphics[width=.45\textwidth]{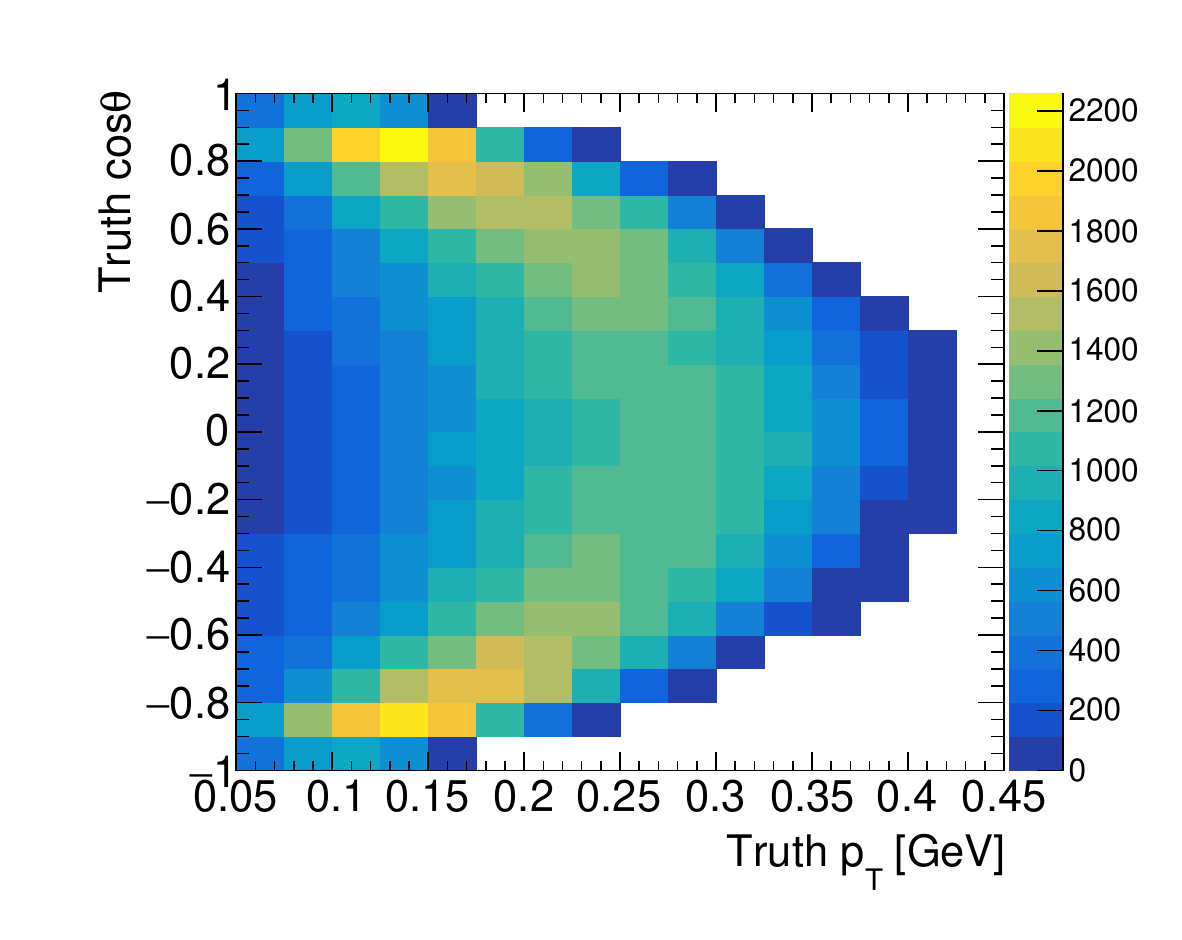} 
\caption{\label{fig:pipijpsi_pt_vs_eta}
The distributions of $p_T$ versus $cos\theta$ for $\mu$ (left) and $\pi$ (right) generated in $\psi(3686)\rightarrow \pi^+\pi^-J/\psi$, $J/\psi\rightarrow \mu^+ \mu^-$ events.
}
\end{figure}

Both single particle ($\mu^-$ and $\pi^-$) events generated using particle gun, and $e^+e^-\rightarrow \psi(3686)\rightarrow \pi^+\pi^-J/\psi, J/\psi\rightarrow \mu^+\mu^-$ events generated with KKMC generator, are used for the tracking performance studies. Each single particle sample is generated with polar angle, $\theta$, and transverse momentum of the particle fixed, and azimuthal angle uniformly distributed from [-$\pi$, $\pi$]. The 2-dimensional distributions of $p_T$ versus $cos\theta$ for $\mu$ and $\pi$ generated in the $\psi(3686)\rightarrow \pi^+\pi^-J/\psi$ events are shown in Figure~\ref{fig:pipijpsi_pt_vs_eta}. The $p_T$ of $\mu$ is in the range of [0.4, 1.8] GeV while that of $\pi$ is in the range of [50, 450] MeV. The Geant4 in OSCAR is used to simulate the hits of the generated final state particles with STCF tracking system immersed in a uniform magnetic field of 1 T.

The detector measurements are created by smearing the position of the simulated hits with Gaussian functions with zero means and widths equivalent to the resolutions of the detectors. For ITK, the hit resolution of 100 $\mu$m  for $l_0$ and 400 $\mu$m for $l_1$ are assumed. For MDC, the resolution of 125 $\mu$m for $l_0$ is assumed.

The tracks are reconstructed as described in Section~\ref{sec:reco}. For a track seed, its transverse momentum is required to be larger than 40 MeV and its transverse impact track parameter is required to be no larger than 10 mm. The $\chi_{min}^2$ = 30 and $n_{\textrm{max}}^{\textrm{branch}} = 1$ for the CKF are used to associate the measurements to tracks. 
The reconstructed tracks are required to have at least 5 measurements on the track. The dependence of the resolution of the fitted perigee track parameters on $p_T$ and $\theta$ of tracks in the acceptance region of the STCF detector is studied using the single particle samples. The track finding performance is studied using the $\psi(3686)\rightarrow \pi^+\pi^-J/\psi$ sample.

\subsection{Track parameters resolution}
The resolution of a track parameter is obtained by fitting the distribution of residuals of the track parameter with a Gaussian function.

\begin{figure}[htbp]
\centering
\includegraphics[width=.45\textwidth]{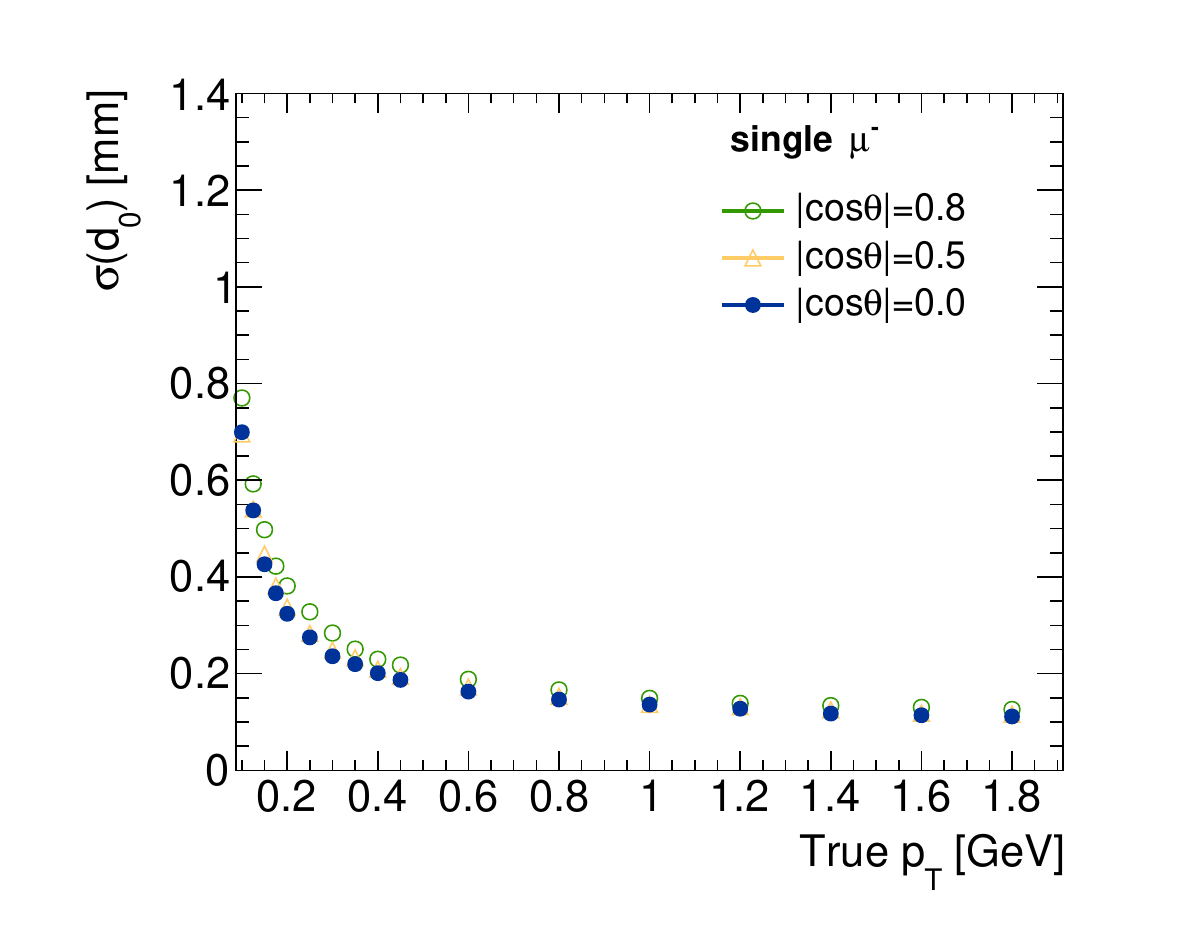}
\includegraphics[width=.45\textwidth]{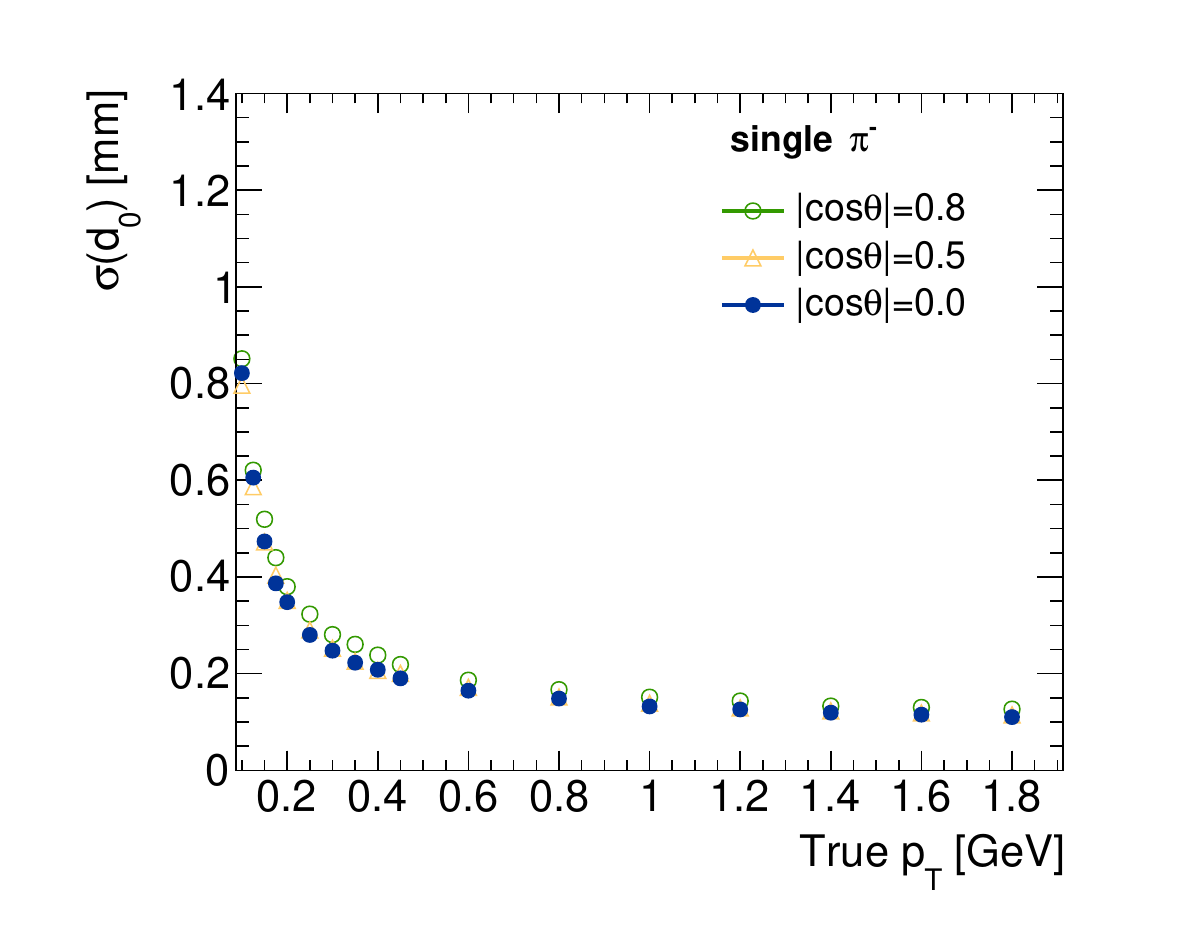} \\
\includegraphics[width=.45\textwidth]{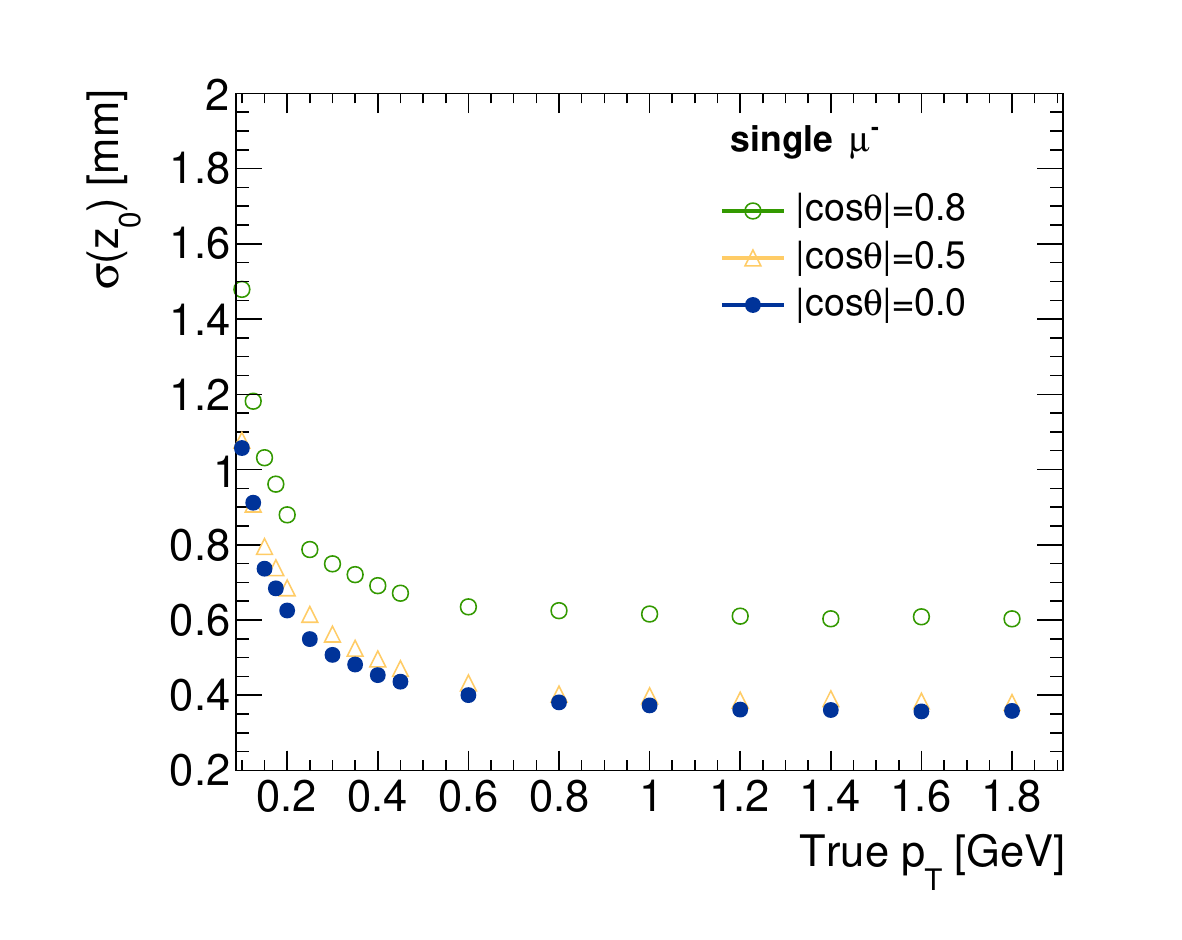}
\includegraphics[width=.45\textwidth]{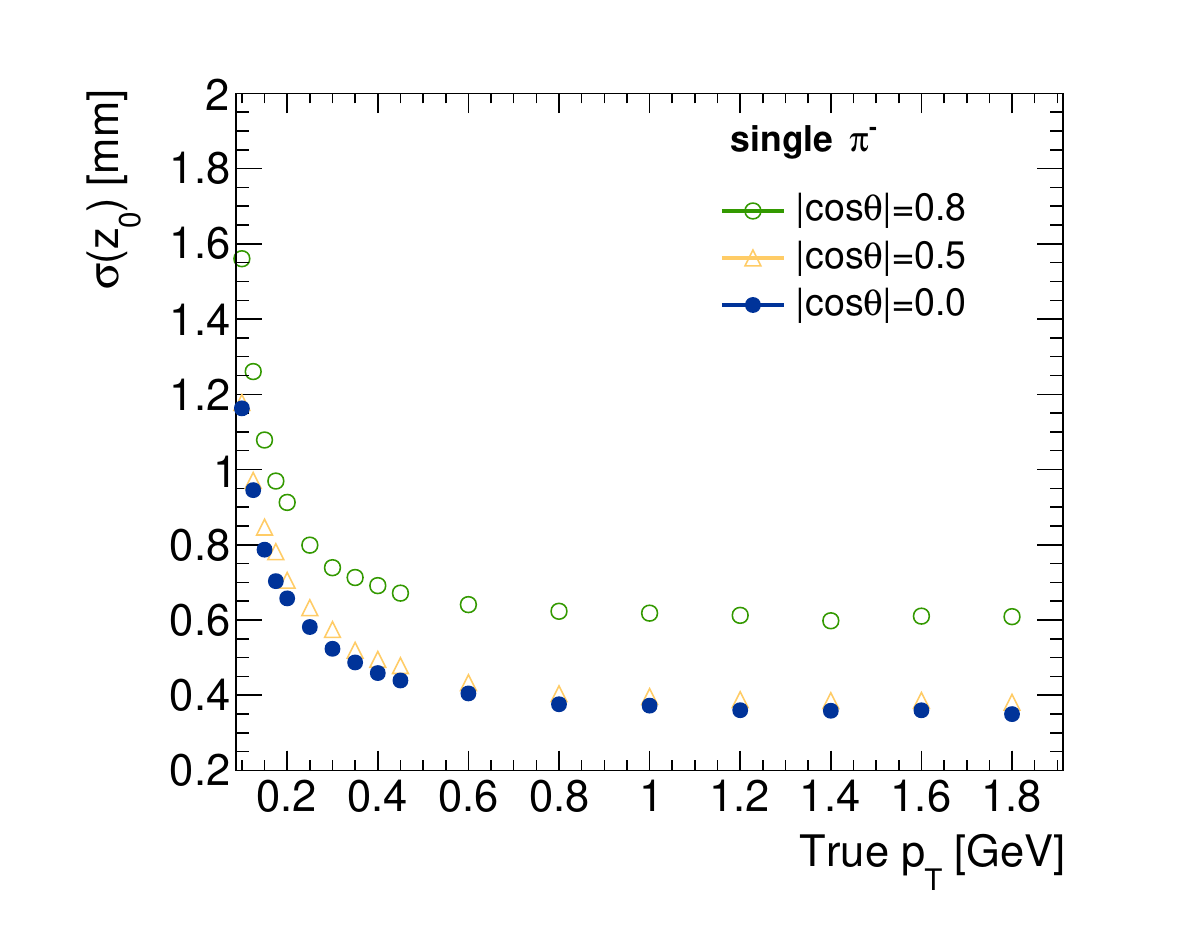} \\
\includegraphics[width=.45\textwidth]{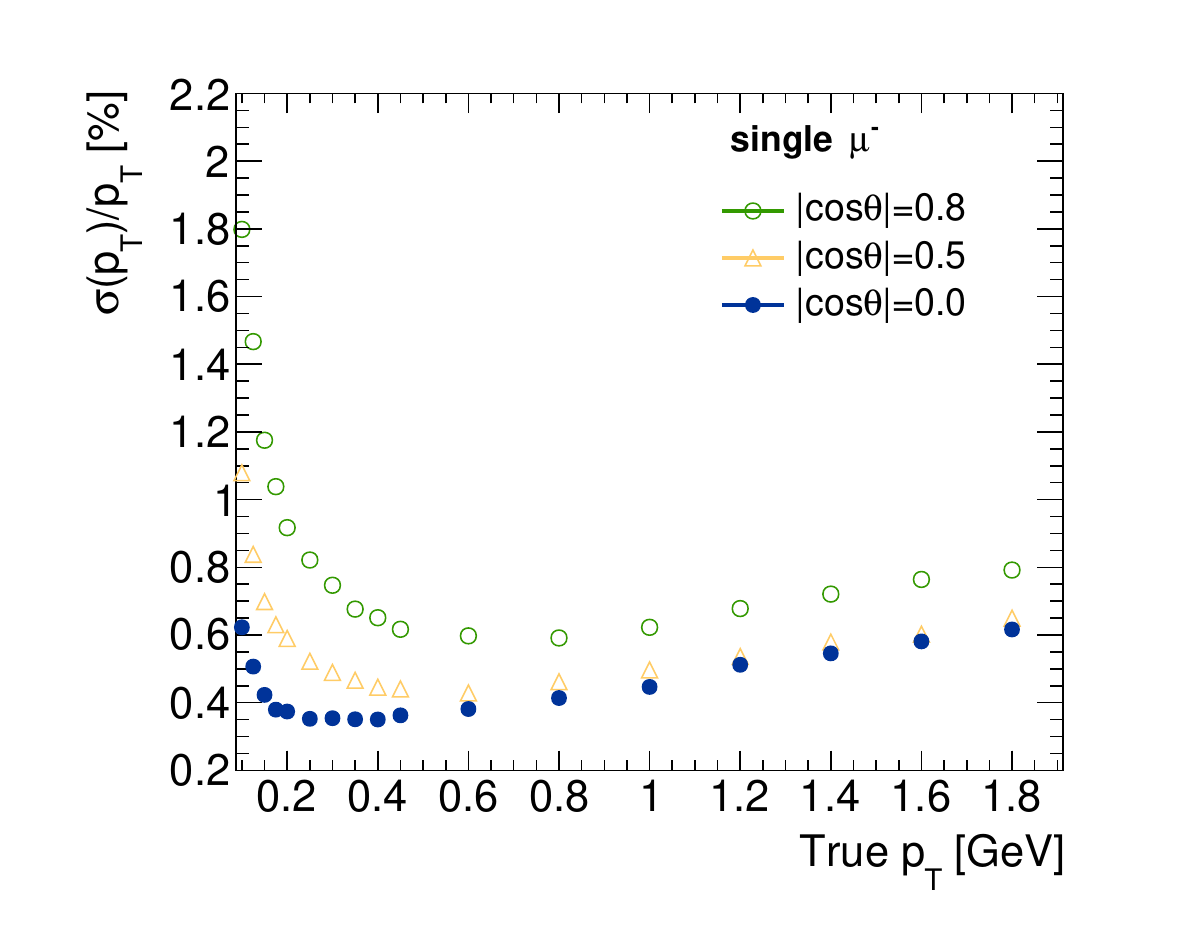}
\includegraphics[width=.45\textwidth]{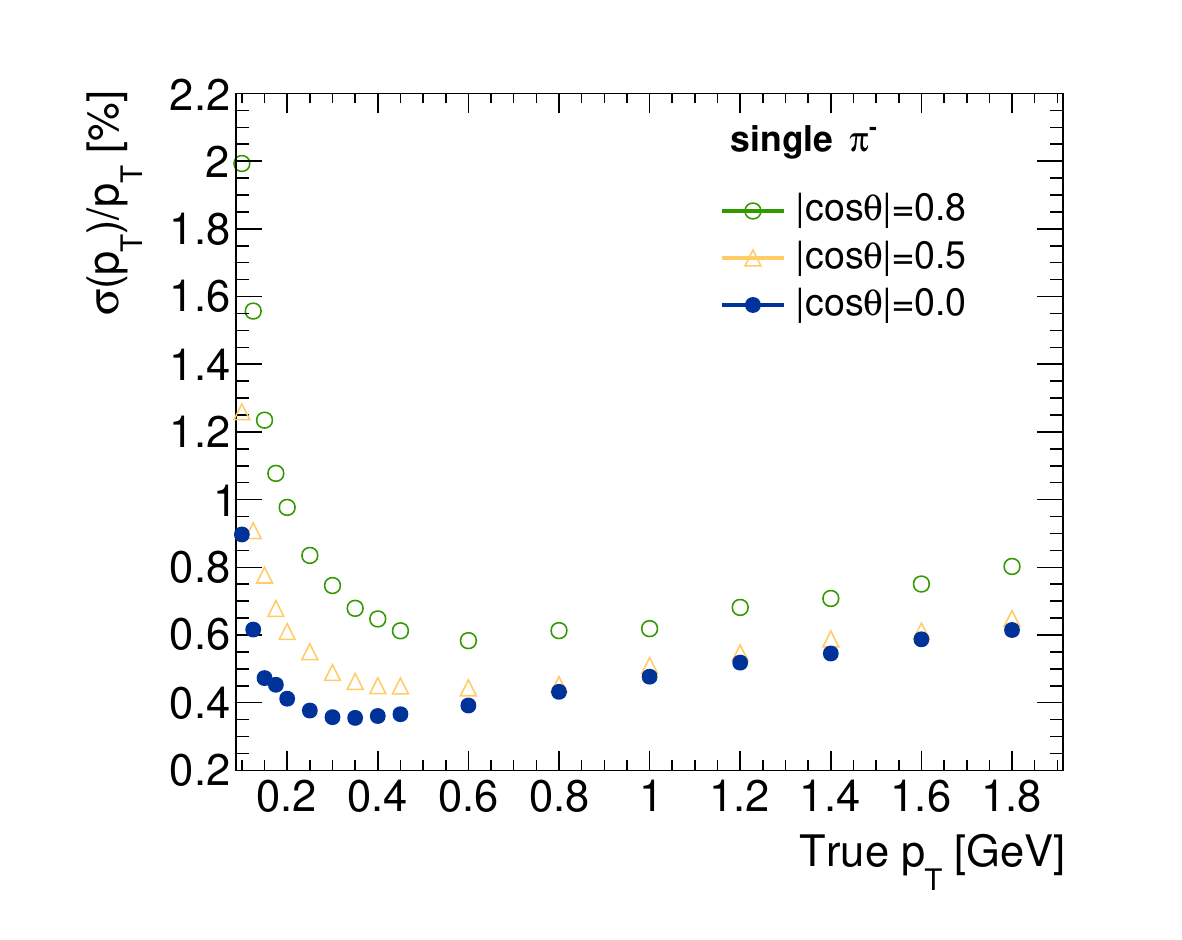}
\caption{\label{fig:resolution_single} The resolution of $d_0$ (top panels), $z_0$ (middle panels) and relative resolution of $p_T$ (bottom panels) for single $\mu^-$ (left panels) and single $\pi^-$ (right panels) as a function of particle $p_T$. The blue dot, yellow triangle and green circle represent the results with $|\textrm{cos}\theta|$ = 0.0, 0.5 and 0.8, respectively. For each $p_T$ and $|\textrm{cos}\theta|$, a sample of 5k events is generated for the study.}
\end{figure}

Figure~\ref{fig:resolution_single} shows the resolution of impact track parameters $d_0$, $z_0$ and transverse momentum $p_T$ as a function of particle $p_T$ at three different polar angles, $|\textrm{cos}\theta|$ = 0.0, 0.5 and 0.8, for single $\mu^-$ and $\pi^-$ events. For $\mu^-$ and $\pi^-$ tracks with $p_T$ = 1 GeV and $\textrm{cos}\theta$ = 0.0, the resolution of $d_0$, $z_0$ and relative resolution of $p_T$ are about 150 $\mu$m, 400 $\mu$m and 0.45\% respectively.

Due to more material effects at lower momentum and larger $|\textrm{cos}\theta|$, the $d_0$ and $z_0$ have a worse resolution at lower $p_T$ and larger $|\textrm{cos}\theta|$. The relative resolution of $p_T$ is also dependent on $p_T$ since the curvature of the track on $x$-$y$ plane is dependent on $p_T$ as well as the material effects. At low momentum range, the material effects are dominant hence the resolution is worse at lower $p_T$. At high momentum range, the resolution is worse with larger $p_T$.

\begin{figure}[htbp]
\centering
\includegraphics[width=.45\textwidth]{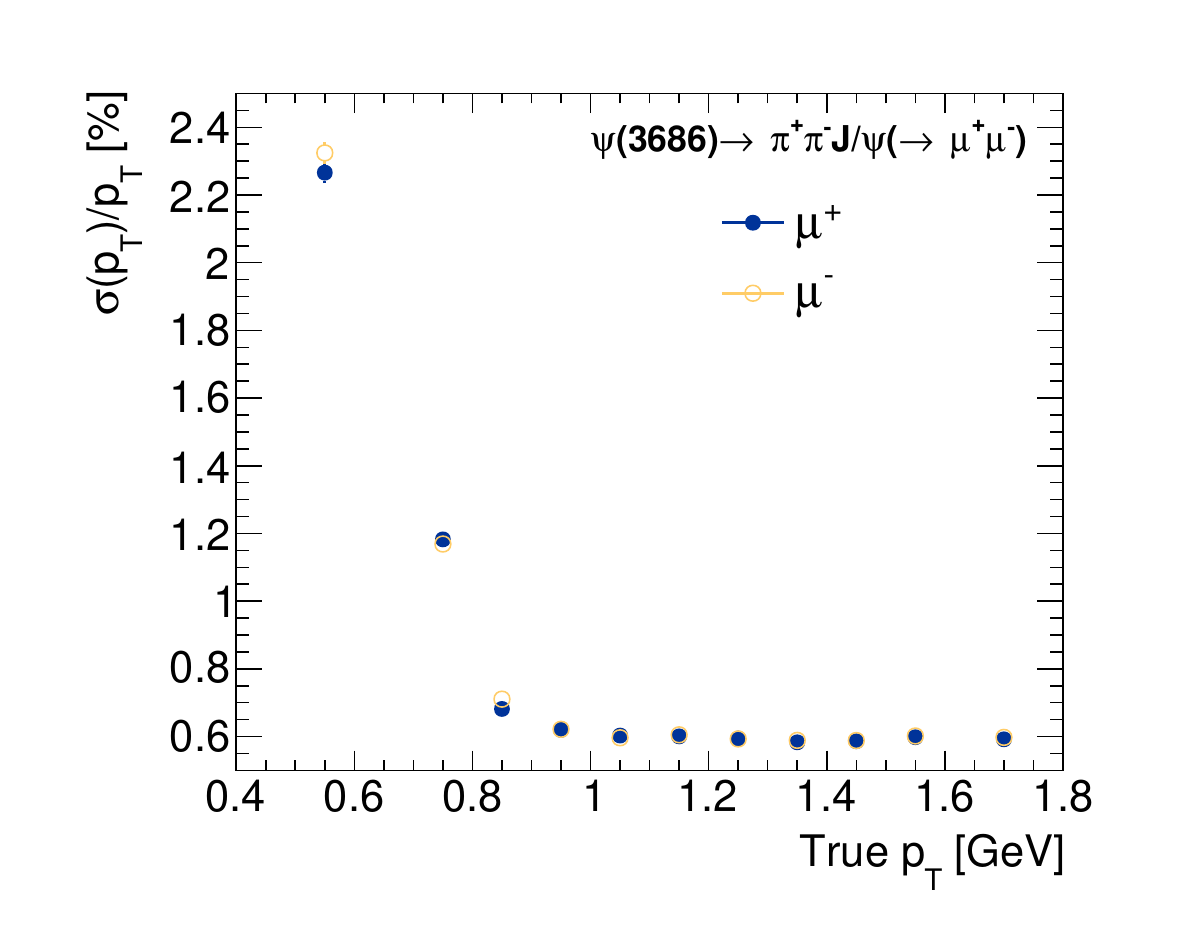}
\includegraphics[width=.45\textwidth]{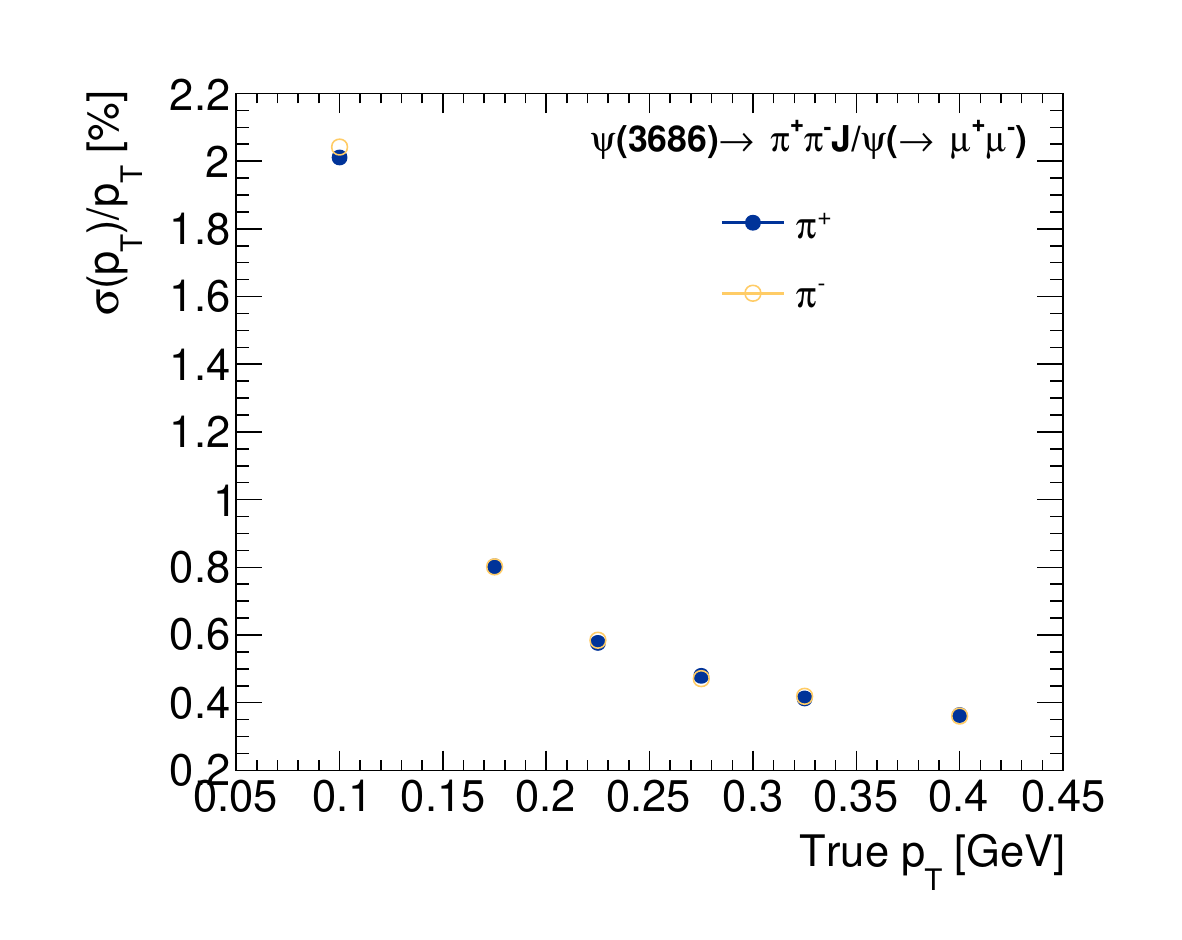} 
\caption{\label{fig:resolution_pipijpsi}The relative resolution of $p_T$ (bottom panels) for $\mu$ (left panels) and $\pi$ (right panels) in $\psi(3686)\rightarrow \pi^+\pi^-J/\psi$, $J/\psi\rightarrow \mu^+ \mu^-$ events as a function of particle $p_T$. The blue dot, and yellow circle represent the results for positive charge particles and negative charge particles, respectively. A sample of 100k events are used.}
\end{figure}

Figure~\ref{fig:resolution_pipijpsi} shows the relative resolution of $p_T$ as a function of particle $p_T$ for $\mu$ and $\pi$ in $\psi(3686)\rightarrow \pi^+\pi^-J/\psi$, $J/\psi\rightarrow \mu^+ \mu^-$ events, where the resolution is related with phase space of the particular physics process.

\subsection{Track finding performance}
The merits often used to characterize the performance of track finding are track reconstruction efficiency, the rate of fake tracks and the rate of duplicate tracks. The classification of a reconstructed track to be fake or duplicate is performed by identifying the simulated particle which has the most simulated hits contributing to this track~\cite{Ai2022,cats}, labelled as \textit{primary particle}. In this study, a reconstructed track is matched to its primary particle if the fraction of its hits from its primary particle, denoted as track purity, is no less than 0.5. A reconstructed track is classified as a fake track if it's not matched to its primary particle. If more than one reconstructed tracks are matched to the same simulated particle, the track with the highest track purity is classified as the real track and others are classified as duplicate tracks. The track reconstruction efficiency is defined by the fraction of particles which have matched reconstructed tracks among the particles which have at least 5 simulated hits. The fake rate is defined by the fraction of fake tracks among the reconstructed tracks. The duplicate rate is defined by the fraction of duplicate tracks among the non-fake reconstructed tracks.

\begin{figure}[htbp]
\centering
\includegraphics[width=.45\textwidth]{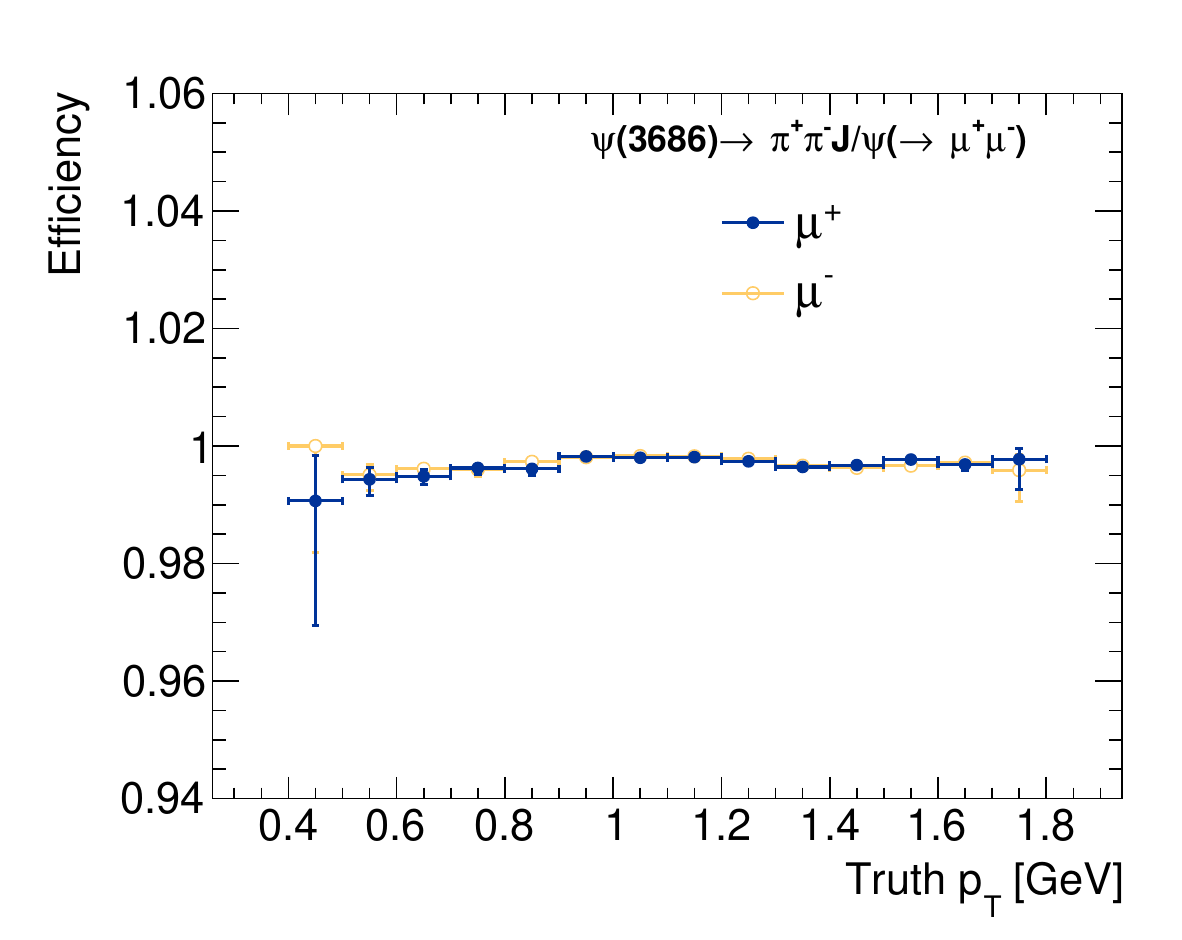}
\includegraphics[width=.45\textwidth]{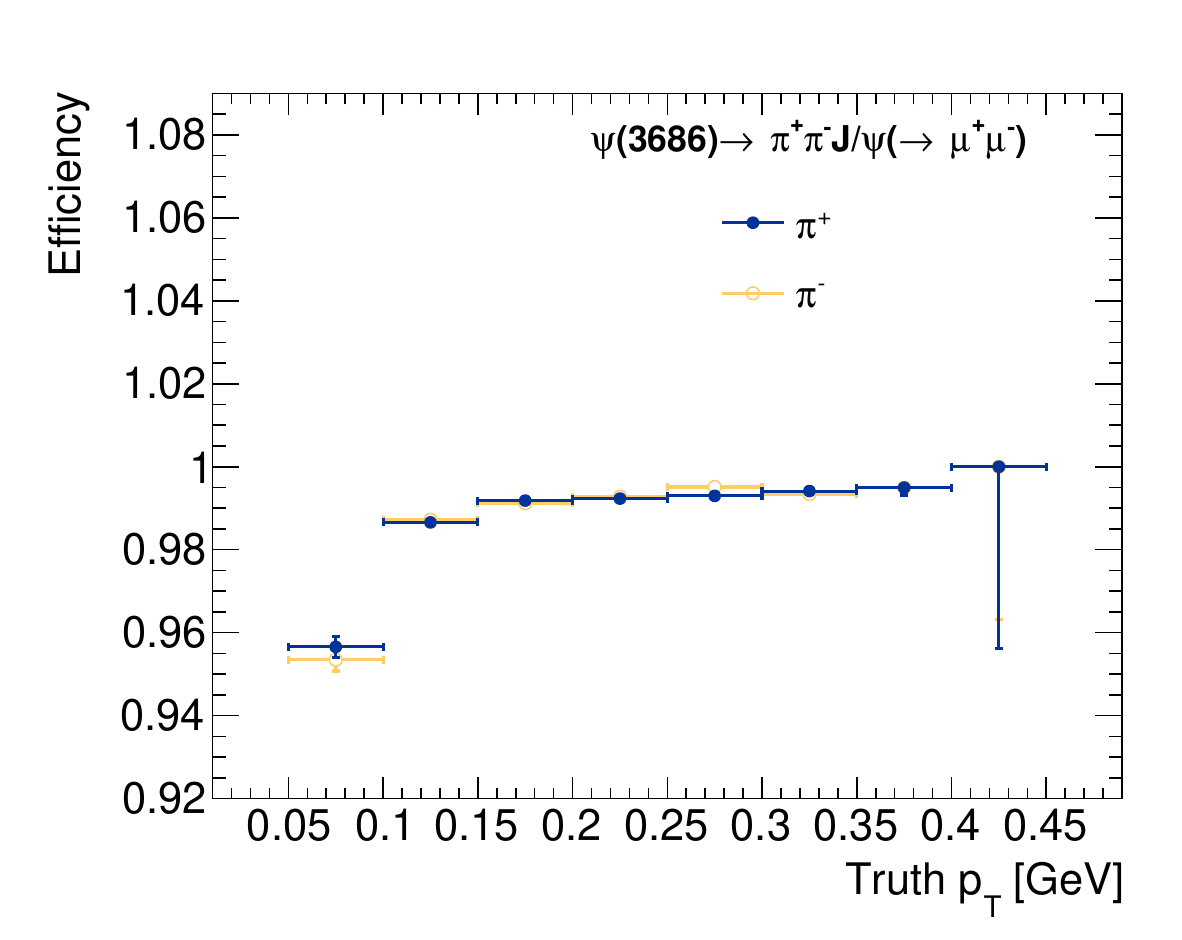} \\
\includegraphics[width=.45\textwidth]{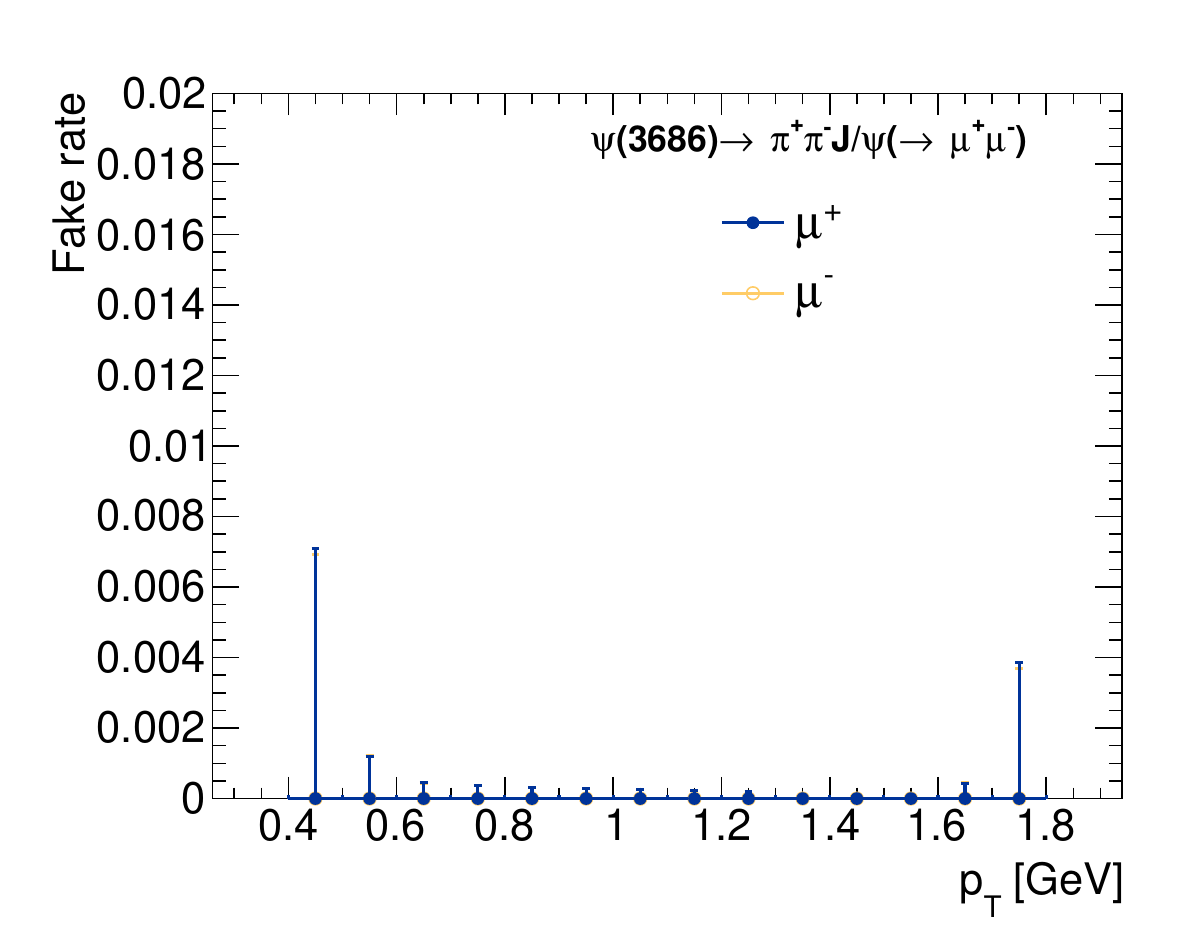}
\includegraphics[width=.45\textwidth]{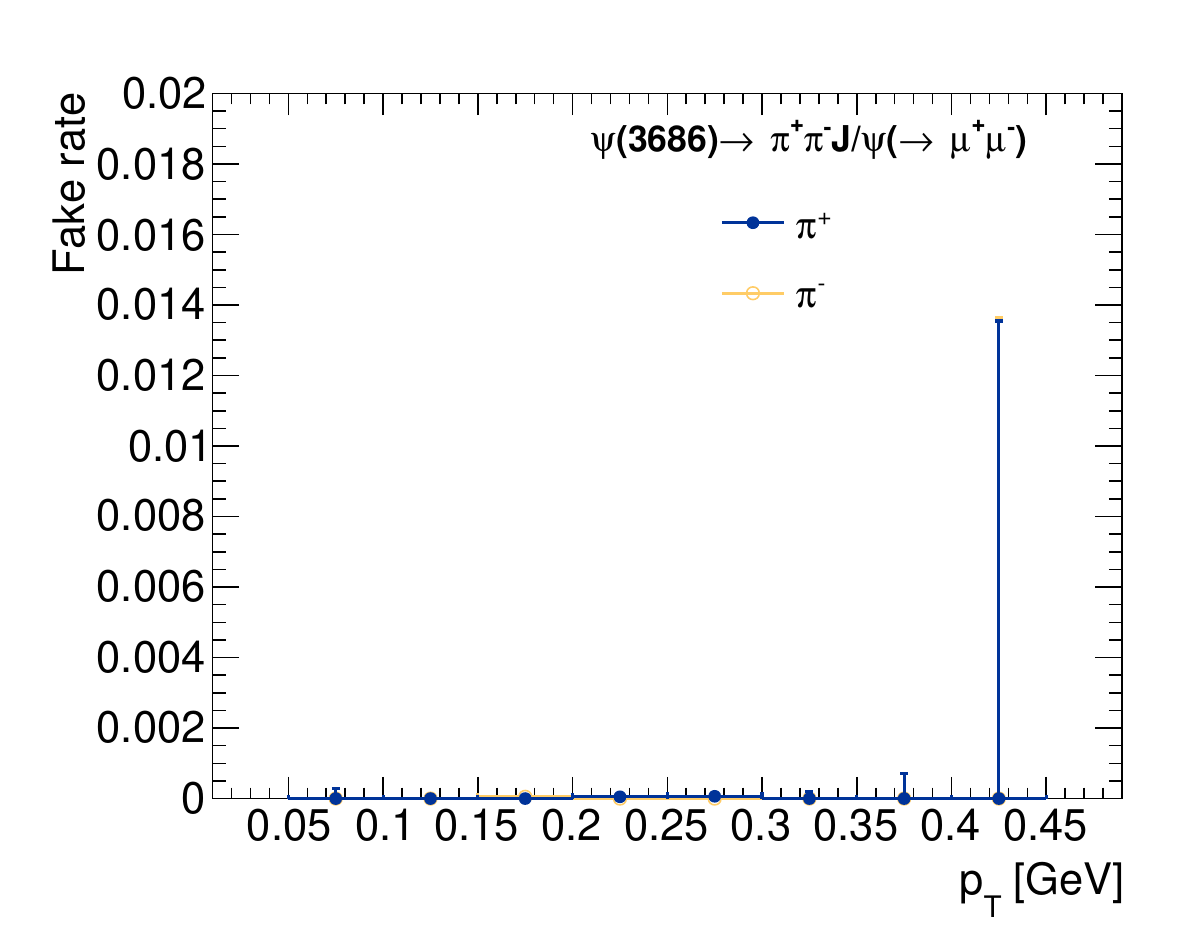} \\
\includegraphics[width=.45\textwidth]{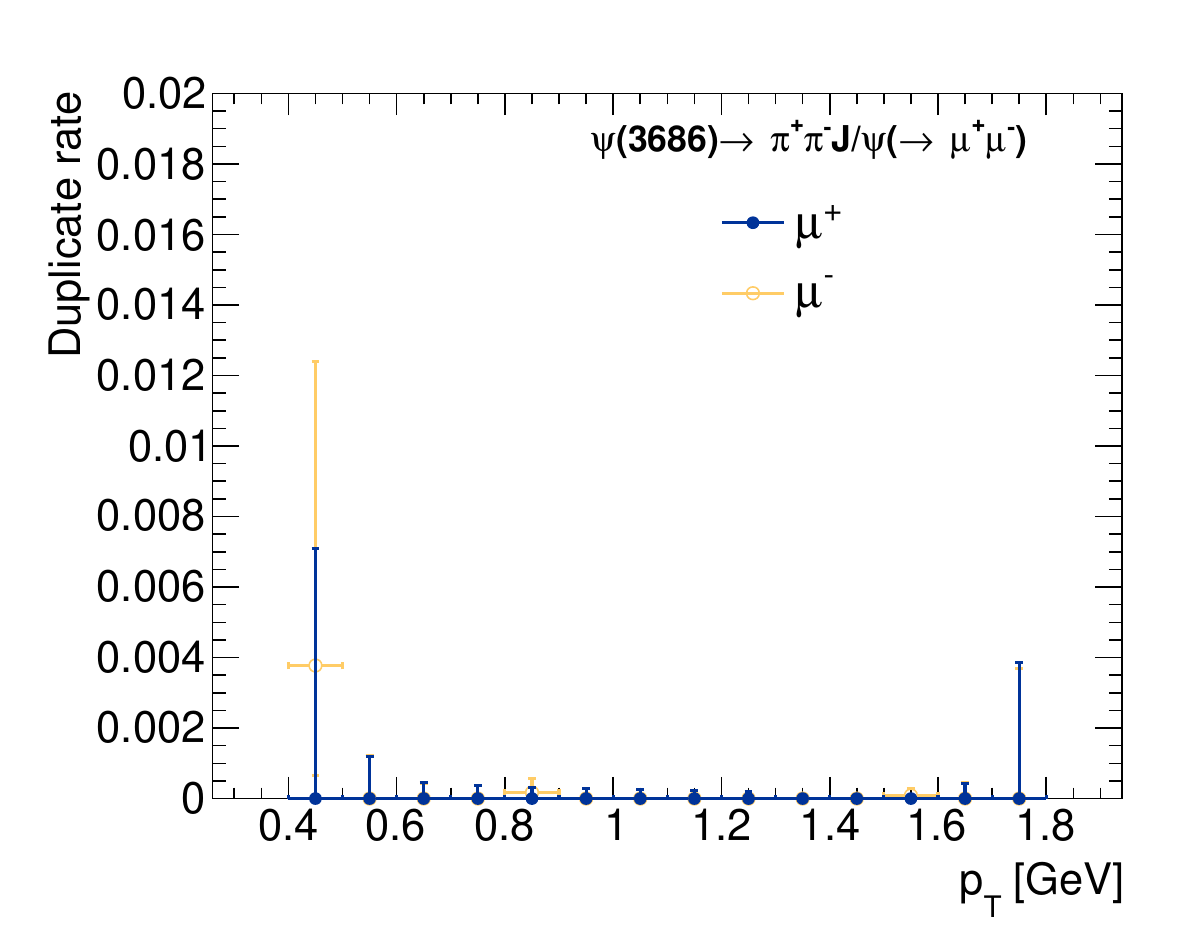}
\includegraphics[width=.45\textwidth]{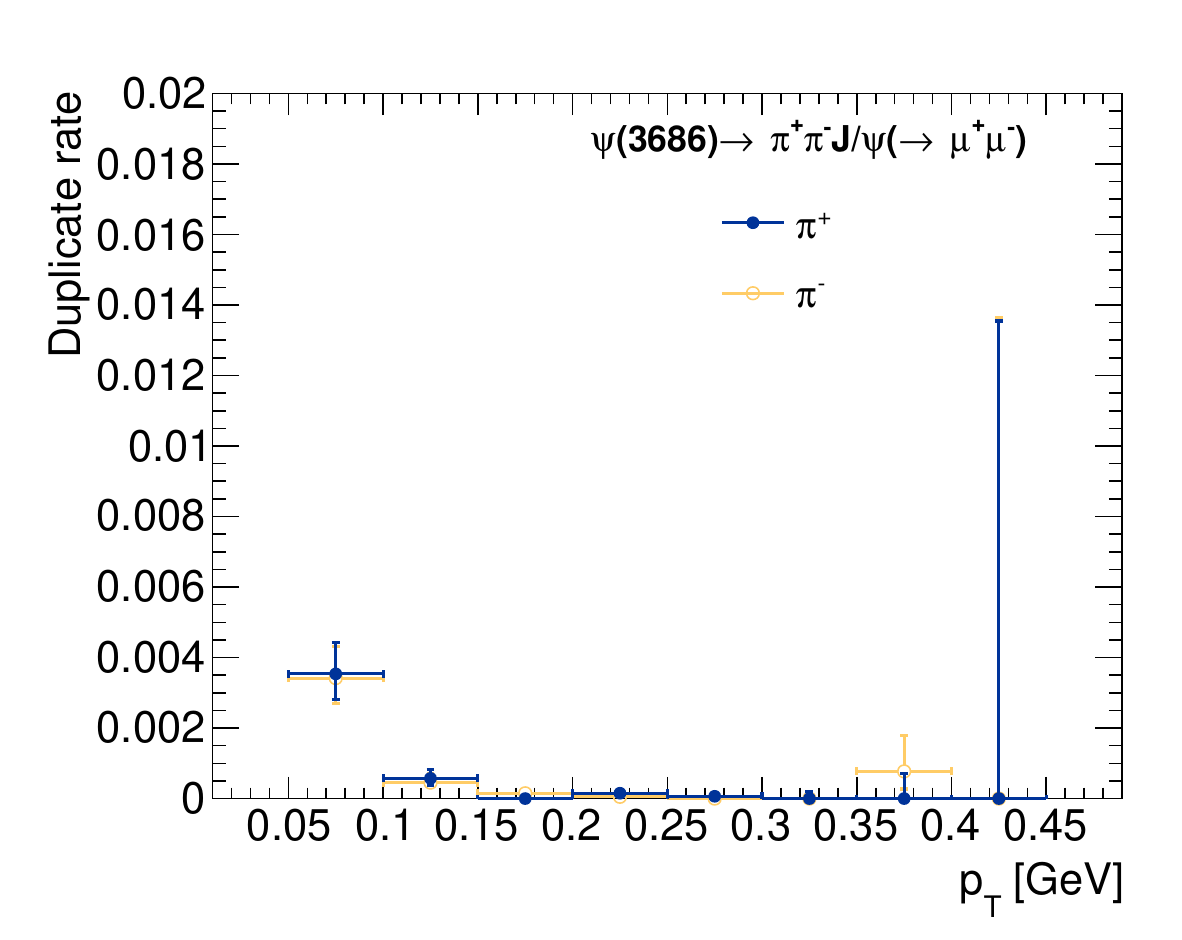} \\
\includegraphics[width=.45\textwidth]{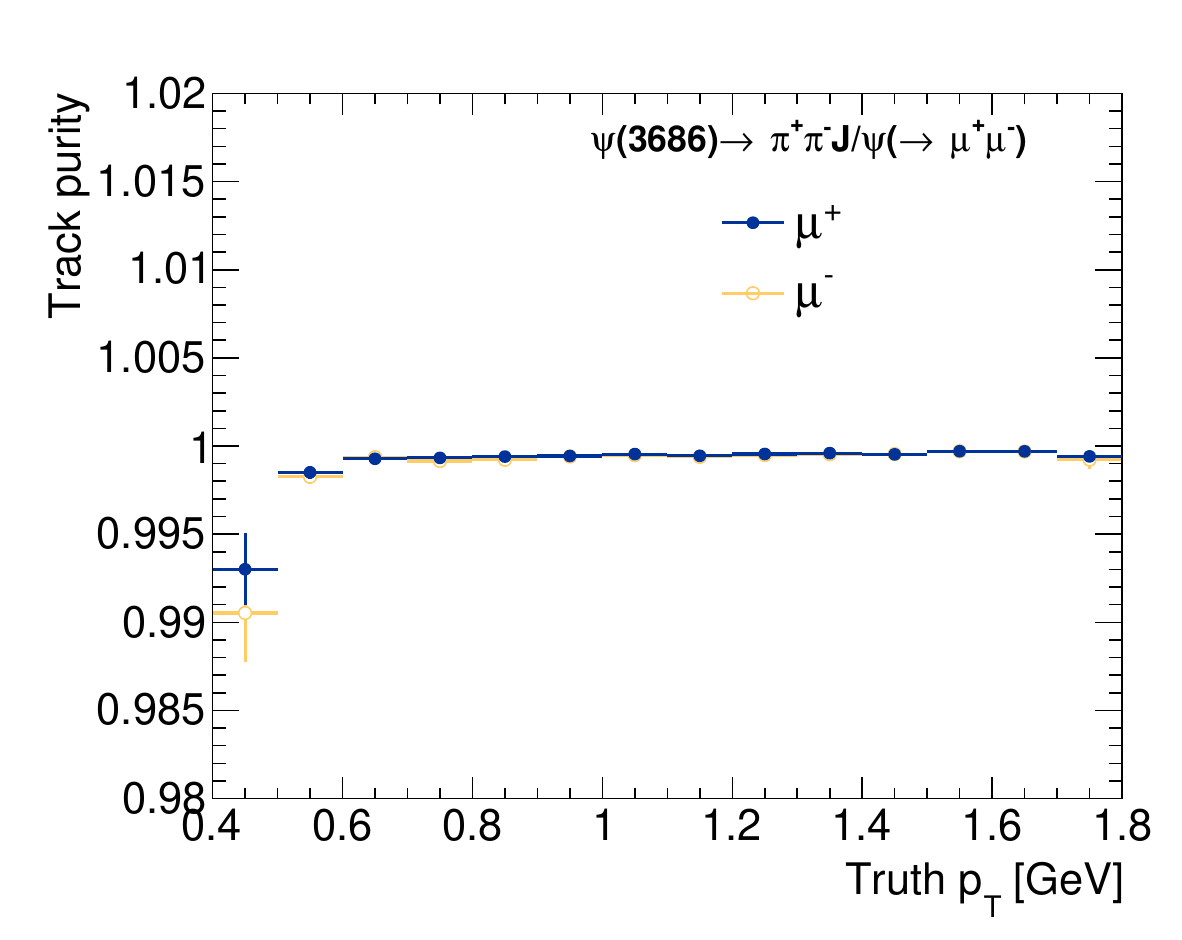}
\includegraphics[width=.45\textwidth]{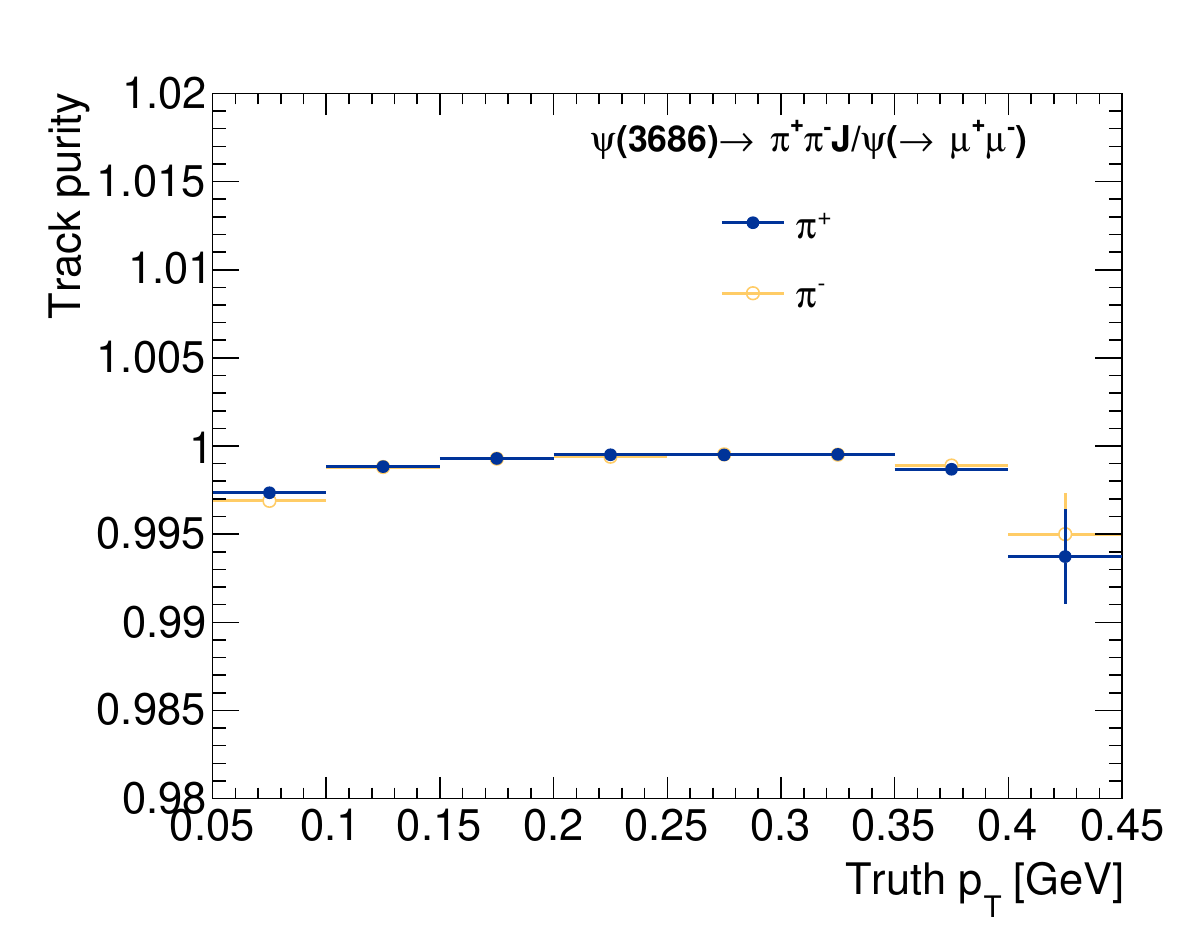} \\
\caption{\label{fig:track_find_perf} The tracking efficiency (first-row panels), fake rate (second-row panels), duplicate rate (third-row panels) and track purity (forth-row panels) for $\mu$ (left panels) and $\pi$ (right panels) with 100k $\psi(3686)\rightarrow \pi^+\pi^-J/\psi$, $J/\psi\rightarrow \mu^+ \mu^-$ events as a function of $p_T$. The blue dot, and yellow circle represent the results for positive charge particles and negative charge particles, respectively.}
\end{figure}

Figure~\ref{fig:track_find_perf} shows the tracking efficiency, fake rate, duplicate rate and track purity for $\mu$ and $\pi$ in $\psi(3686)\rightarrow \pi^+\pi^-J/\psi$ events. For $\mu$ and $\pi$ with $p_T$ above 150 MeV, the tracking efficiency is above 99\%. For $\pi$ with momentum in the range of [50, 100] MeV, a tracking efficiency of 95\% is achieved. The fake rate is found to be negligible. For $\pi$ track with $p_T$ below 150 MeV, duplicate tracks due to looping trajectories of the particles are found with the duplicate rate below 0.4\%.

\section{Conclusion}
\label{sec:conclusion}

To achieve the physics goals of the STCF physics program, which is dedicated to studies of hadron physics, the asymmetry of matter-antimatter and searches for physics beyond SM in the $\tau$-charm sector in the post-BEPCII era, the tracking system of STCF is required to provide sufficient performance for charged tracks reconstruction.  

We implemented the common tracking toolkit ACTS for track reconstruction at STCF based on a three-layer $\mu$RWELL-based ITK and a 48-layer MDC, and demonstrated the performance of ACTS for a drift chamber for the first time. The performance of track reconstruction was studied on top of Geant4-based full simulation of interaction of particles with detectors. The track finding and track fitting was performed based on the CKF in ACTS. In this study where backgrounds were not considered yet, the CKF was configured to run in a sequential mode with only the best measurement associated to the track at each propagation, which was found to result in the best overall tracking performance. The tracking efficiency is above 95\% for tracks with $p_T$ above 50 MeV. For tracks with $p_T > $ 150 MeV, a tracking efficiency above 99\% is achieved. The rate of fake tracks is negligible and the rate of duplicate tracks, which arise in the range of $p_T<$ 150 MeV, is below 0.4\%. The $\sigma(p_T)/p_T$ is below 0.5\% when $p_T = $ 1 GeV and |cos$\theta$| $\leq$ 0.5. 
Those excellent performance shows that ACTS is a very promising tracking toolkit to be used for further optimization of the design and geometry layout of the STCF tracking system. Meanwhile, track finding using Hough Transform for STCF is being studied. The tracking performance considering beam induced background will be explored in future studies.

\acknowledgments
The authors are grateful to the STCF group for the profitable discussions and express gratitude to the Hefei Comprehensive National Science Center for their strong support. This work is supported by the international partnership program of the Chinese Academy of Sciences (Grant No. 211134KYSB20200057) and the National Natural Science Foundation of China (Grant No. 12025502).




\bibliography{bibfile}

\begin{thebibliography}{10}
\expandafter\ifx\csname url\endcsname\relax
  \def\url#1{\texttt{#1}}\fi
\expandafter\ifx\csname urlprefix\endcsname\relax\def\urlprefix{URL }\fi
\expandafter\ifx\csname href\endcsname\relax
  \def\href#1#2{#2} \def\path#1{#1}\fi

\bibitem{ABLIKIM2010345}
M.~Ablikim, et~al., {Design and construction of the BESIII detector}, Nucl.
  Instrum. Meth. A 614~(3) (2010) 345--399.
\newblock \href {https://doi.org/https://doi.org/10.1016/j.nima.2009.12.050}
  {\path{doi:https://doi.org/10.1016/j.nima.2009.12.050}}.

\bibitem{Luo:IPAC2019-MOPRB031}
Q.~Luo, W.~Gao, J.~Lan, W.~Li, D.~Xu, {P}rogress of {C}onceptual {S}tudy for
  the {A}ccelerators of a 2{-}7 {G}e{V} {S}uper {T}au {C}harm {F}acility at
  {C}hina, in: Proc. 10th International Particle Accelerator Conference
  (IPAC'19), Melbourne, Australia, 19-24 May 2019, no.~10 in International
  Particle Accelerator Conference, JACoW Publishing, Geneva, Switzerland, 2019,
  pp. 643--645, https://doi.org/10.18429/JACoW-IPAC2019-MOPRB031.
\newblock \href {https://doi.org/doi:10.18429/JACoW-IPAC2019-MOPRB031}
  {\path{doi:doi:10.18429/JACoW-IPAC2019-MOPRB031}}.

\bibitem{Ai2022}
X.~Ai, C.~Allaire, N.~Calace, A.~Czirkos, M.~Elsing, I.~Ene, R.~Farkas, L.-G.
  Gagnon, R.~Garg, P.~Gessinger, H.~Grasland, H.~M. Gray, C.~Gumpert,
  J.~Hrdinka, B.~Huth, M.~Kiehn, F.~Klimpel, B.~Kolbinger, A.~Krasznahorkay,
  R.~Langenberg, C.~Leggett, G.~Mania, E.~Moyse, J.~Niermann, J.~D. Osborn,
  D.~Rousseau, A.~Salzburger, B.~Schlag, L.~Tompkins, T.~Yamazaki, B.~Yeo,
  J.~Zhang, {A Common Tracking Software Project}, Computing and Software for
  Big Science 6~(1) (2022) 8.
\newblock \href {https://doi.org/10.1007/s41781-021-00078-8}
  {\path{doi:10.1007/s41781-021-00078-8}}.

\bibitem{Salzburger_A_Common_Tracking_2021}
A.~Salzburger, P.~Gessinger, F.~Klimpel, M.~Kiehn, B.~Schlag, H.~Grasland,
  R.~Langenberg, C.~Allaire, X.~Ai, B.~Huth, L.-G. Gagnon, N.~Calace,
  A.~Krasznahorkay, B.~Yeo, C.~Leggett, C.~Angela, T.~Yamazaki, J.~Niermann,
  I.~Ene, J.~Osborn, R.~Bala~Garg, A.~Stefl, L.~F. Coelho, A.~J. Pfleger,
  G.~Almeida, S.~N. Swatman, \href{https://github.com/acts-project/acts}{{A
  Common Tracking Software Project}} (7 2021).
\newblock \href {https://doi.org/10.5281/zenodo.5141419}
  {\path{doi:10.5281/zenodo.5141419}}.
\newline\urlprefix\url{https://github.com/acts-project/acts}

\bibitem{ATL-PHYS-PUB-2021-012}
{ATLAS Collaboration}, {Software Performance of the ATLAS Track Reconstruction
  for LHC Run 3}, Tech. rep., CERN, Geneva, all figures including auxiliary
  figures are available at
  https://atlas.web.cern.ch/Atlas/GROUPS/PHYSICS/PUBNOTES/ATL-PHYS-PUB-2021-012
  (May 2021).

\bibitem{Osborn2021}
J.~D. Osborn, A.~D. Frawley, J.~Huang, S.~Lee, H.~P.~D. Costa, M.~Peters,
  C.~Pinkenburg, C.~Roland, H.~Yu, {Implementation of ACTS into sPHENIX Track
  Reconstruction}, Computing and Software for Big Science 5~(1) (2021) 23.
\newblock \href {https://doi.org/10.1007/s41781-021-00068-w}
  {\path{doi:10.1007/s41781-021-00068-w}}.

\bibitem{Bencivenni_2017}
G.~Bencivenni, L.~Benussi, L.~Borgonovi, R.~de~Oliveira, P.~D. Simone,
  G.~Felici, M.~Gatta, P.~Giacomelli, G.~Morello, A.~Ochi, M.~P. Lener,
  A.~Ranieri, M.~Ressegotti, E.~Tskhadadze, I.~Vai, V.~Valentino, {The
  $\mu$-RWELL detector}, Journal of Instrumentation 12~(06) (2017) C06027.
\newblock \href {https://doi.org/10.1088/1748-0221/12/06/C06027}
  {\path{doi:10.1088/1748-0221/12/06/C06027}}.

\bibitem{kalman1960}
R.~E. Kalman, {A New Approach to Linear Filtering and Prediction Problems},
  Journal of Basic Engineering 82~(1) (1960) 35--45.
\newblock \href {https://doi.org/10.1115/1.3662552}
  {\path{doi:10.1115/1.3662552}}.

\bibitem{ADAM2005281}
I.~Adam, et~al., {The DIRC particle identification system for the BaBar
  experiment}, Nucl. Instrum. Meth. A 538~(1) (2005) 281--357.
\newblock \href {https://doi.org/https://doi.org/10.1016/j.nima.2004.08.129}
  {\path{doi:https://doi.org/10.1016/j.nima.2004.08.129}}.

\bibitem{oscar}
W.~H. Huang, H.~Li, H.~Zhou, T.~Li, Q.~Y. Li, X.~T. Huang, {Design and
  Development of the Core Software for STCF Offline Data Processing} (2022).
\newblock \href {https://doi.org/10.48550/ARXIV.2211.03137}
  {\path{doi:10.48550/ARXIV.2211.03137}}.

\bibitem{kkmc}
S.~Jadach, B.~F.~L. Ward, Z.~Wa\ifmmode~\mbox{\c{}}\else \c{}\fi{}s, {Coherent
  exclusive exponentiation for precision Monte Carlo calculations}, Phys. Rev.
  D 63 (2001) 113009.
\newblock \href {https://doi.org/10.1103/PhysRevD.63.113009}
  {\path{doi:10.1103/PhysRevD.63.113009}}.

\bibitem{evtgen}
D.~J. Lange, {The EvtGen particle decay simulation package}, Nucl. Instrum.
  Meth. A 462~(1) (2001) 152--155, bEAUTY2000, Proceedings of the 7th Int.
  Conf. on B-Physics at Hadron Machines.
\newblock \href {https://doi.org/https://doi.org/10.1016/S0168-9002(01)00089-4}
  {\path{doi:https://doi.org/10.1016/S0168-9002(01)00089-4}}.

\bibitem{dd4hep}
M.~Frank, F.~Gaede, C.~Grefe, P.~Mato, {DD4hep: A Detector Description Toolkit
  for High Energy Physics Experiments}, Journal of Physics: Conference Series
  513~(2) (2014) 022010.
\newblock \href {https://doi.org/10.1088/1742-6596/513/2/022010}
  {\path{doi:10.1088/1742-6596/513/2/022010}}.

\bibitem{xml}
{Extensible Markup Language (XML) webpage}, \url{https://www.w3.org/XML}.

\bibitem{AGOSTINELLI2003250}
S.~Agostinelli, et~al., {Geant4—a simulation toolkit}, Nucl. Instrum. Meth. A
  506~(3) (2003) 250--303.
\newblock \href {https://doi.org/https://doi.org/10.1016/S0168-9002(03)01368-8}
  {\path{doi:https://doi.org/10.1016/S0168-9002(03)01368-8}}.

\bibitem{Fruhwirth2021}
R.~Fr{\"u}hwirth, A.~Strandlie, {Track Finding}, Springer International
  Publishing, Cham, 2021, pp. 81--102.
\newblock \href {https://doi.org/10.1007/978-3-030-65771-0_5}
  {\path{doi:10.1007/978-3-030-65771-0_5}}.

\bibitem{Braun:1540}
N.~Braun, P.~D.~M. Feindt, P.~D. F.~U. Bernlochner, {Combinatorial Kalman
  Filter and High Level Trigger Reconstruction for the Belle II Experiment},
  Ph.D. thesis, Karlsuhe, Karlsuhe Institute of Technology, Karlsruhe,
  presented on 21 12 2018 (2018).

\bibitem{BERTACCHI2021107610}
V.~Bertacchi, T.~Bilka, N.~Braun, G.~Casarosa, L.~Corona, S.~Cunliffe,
  F.~Dattola, G.~{De Marino}, M.~{De Nuccio}, G.~{De Pietro}, T.~{Van Dong},
  G.~Dujany, P.~Ecker, M.~Eliachevitch, T.~Fillinger, O.~Frost, R.~Frühwirth,
  U.~Gebauer, S.~Glazov, N.~Gosling, A.~Guo, T.~Hauth, M.~Heck, M.~Kaleta,
  J.~Kandra, C.~Kleinwort, T.~Kuhr, S.~Kurz, P.~Kvasnicka, J.~Lettenbichler,
  T.~Lueck, A.~Martini, F.~Metzner, D.~Neverov, C.~Niebuhr, E.~Paoloni,
  S.~Patra, L.~Piilonen, C.~Praz, M.~T. Prim, C.~Pulvermacher, S.~Racs, N.~Rad,
  P.~Rados, M.~Ritter, G.~Rizzo, A.~Rostomyan, B.~Scavino, T.~Schlüter,
  B.~Schwenker, S.~Spataro, B.~Spruck, H.~Svidras, F.~Tenchini, Y.~Uematsu,
  J.~Webb, C.~Wessel, L.~Zani, {Track finding at Belle II}, Computer Physics
  Communications 259 (2021) 107610.
\newblock \href {https://doi.org/https://doi.org/10.1016/j.cpc.2020.107610}
  {\path{doi:https://doi.org/10.1016/j.cpc.2020.107610}}.

\bibitem{BRUN2003676}
R.~Brun, A.~Gheata, M.~Gheata, {The ROOT geometry package}, Nucl. Instrum.
  Methods. Phys. Res. A 502~(2) (2003) 676--680.
\newblock \href {https://doi.org/10.1016/S0168-9002(03)00541-2}
  {\path{doi:10.1016/S0168-9002(03)00541-2}}.

\bibitem{Brun:1997pa}
R.~Brun, F.~Rademakers, {ROOT: An object oriented data analysis framework},
  Nucl. Instrum. Meth. A 389 (1997) 81--86.
\newblock \href {https://doi.org/10.1016/S0168-9002(97)00048-X}
  {\path{doi:10.1016/S0168-9002(97)00048-X}}.

\bibitem{oscar_manual}
{OSCAR user manual: Track reconstruction with ACTS toolkit},
  \url{http://202.141.163.203:8008/oscar_manual/reconstruction/trackingWithACTS.html}.

\bibitem{garg2023exploration}
R.~B. Garg, E.~Hofgard, L.~Tompkins, H.~Gray, {Exploration of different
  parameter optimization algorithms within the context of ACTS software
  framework} (2023).
\newblock \href {http://arxiv.org/abs/2211.00764} {\path{arXiv:2211.00764}}.

\bibitem{optuna}
{Optuna HyperParamter Optimization Framework}, \url{https://optuna.org/}.

\bibitem{Duda1972UseOT}
R.~Duda, P.~Hart, {Use of the Hough transformation to detect lines and curves
  in pictures}, Commun. ACM 15 (1972) 11--15.

\bibitem{FRUHWIRTH1999197}
R.~Frühwirth, A.~Strandlie, {Track fitting with ambiguities and noise: A study
  of elastic tracking and nonlinear filters}, Computer Physics Communications
  120~(2) (1999) 197--214.
\newblock \href {https://doi.org/https://doi.org/10.1016/S0010-4655(99)00231-3}
  {\path{doi:https://doi.org/10.1016/S0010-4655(99)00231-3}}.

\bibitem{cats}
I.~Kisel, S.~Masciocchi, {CATS - A Cellular Automaton for Tracking in Silicon
  for the HERA-B Vertex Detector} (01 2000).

\end{thebibliography}







\end{document}